\documentclass{aa}

\usepackage{graphicx}
\usepackage{CJK}
\usepackage{enumitem}
\usepackage[varg]{txfonts}
\usepackage{amsmath}
\usepackage{amssymb}
\usepackage{footmisc}
\usepackage{booktabs}
\usepackage[colorlinks = true,
            linkcolor = blue,
            urlcolor  = blue,
            citecolor = blue,
            anchorcolor = blue,
            unicode]{hyperref}

\begin{document}

\begin{CJK*}{UTF8}{gbsn}

\title{Evolved Massive Stars at Low-metallicity\\ \uppercase\expandafter{\romannumeral1}. A Source Catalog for the Small Magellanic Cloud}
\titlerunning{Evolved Massive Stars at low-Z \uppercase\expandafter{\romannumeral1}. A Source Catalog for the SMC}

\author{
Ming Yang (杨明) \inst{1} \and Alceste Z. Bonanos \inst{1} \and Bi-Wei Jiang (姜碧沩) \inst{2} \and Jian Gao (高健) \inst{2} \and Panagiotis Gavras \inst{3} \and Grigoris Maravelias \inst{1} \and Yi Ren (任逸) \inst{2} \and Shu Wang (王舒) \inst{4} \and Meng-Yao Xue (薛梦瑶) \inst{5} \and Frank Tramper \inst{1} \and Zoi T. Spetsieri \inst{1,6} \and Ektoras Pouliasis \inst{1,6}
}
\authorrunning{Yang, Bonanos \& Jiang et al.}

\institute{
IAASARS, National Observatory of Athens, Vas. Pavlou and I. Metaxa, Penteli 15236, Greece\\
                \email{myang@noa.gr} \and
Department of Astronomy, Beijing Normal University, Beijing 100875, People's Republic of China \and
Rhea Group for ESA/ESAC, Camino bajo del Castillo, s/n, Urbanizacion Villafranca del Castillo, Villanueva de la Cañada, 28692 Madrid, Spain \and
Key Laboratory of Optical Astronomy, National Astronomical Observatories, Chinese Academy of Sciences, Datun Road 20A, Beijing 100101, People's Republic of China \and
International Centre for Radio Astronomy Research, Curtin University, Bentley, WA 6102, Australia \and
Department of Astrophysics, Astronomy \& Mechanics, Faculty of Physics, University of Athens, Zografos, 15783 Athens, Greece
}

\abstract{
We present a clean, magnitude-limited (IRAC1 or WISE1 $\leq$ 15.0 mag) multiwavelength source catalog for the Small Magellanic Cloud (SMC) with 45,466 targets in total, with the purpose of building an anchor for future studies, especially for the massive star populations at low-metallicity. The catalog contains data in 50 different bands including 21 optical and 29 infrared bands, retrieved from SEIP, VMC, IRSF, \textit{AKARI}, HERITAGE, \textit{Gaia}, SkyMapper, NSC, Massey et al. (2002), and \textit{GALEX}, ranging from the ultraviolet to the far-infrared. Additionally, radial velocities and spectral classifications were collected from the literature, as well as infrared and optical variability statistics were retrieved from \textit{WISE}, SAGE-Var, VMC, IRSF, \textit{Gaia}, NSC, and OGLE. The catalog was essentially built upon a $1''$ crossmatching and a $3''$ deblending between the \textit{Spitzer} Enhanced Imaging Products (SEIP) source list and \textit{Gaia} Data Release 2 (DR2) photometric data. Further constraints on the proper motions and parallaxes from \textit{Gaia} DR2 allowed us to remove the foreground contamination. We estimated that about 99.5\% of the targets in our catalog were most likely genuine members of the SMC. By using the evolutionary tracks and synthetic photometry from MESA Isochrones \& Stellar Tracks and the theoretical $\rm J-K_S$ color cuts, we identified 1,405 red supergiant, 217 yellow supergiant and 1,369 blue supergiant candidates in the SMC in five different color-magnitude diagrams (CMDs), where attention should also be paid to the incompleteness of our sample. We ranked the candidates based on the intersection of different CMDs. A comparison between the models and observational data shows that the lower limit of initial mass for the RSGs population may be as low as 7 or even 6 $M_{\sun}$ and the RSG is well separated from the AGB population even at faint magnitude, making RSGs a unique population connecting the evolved massive and intermediate stars, since stars with initial mass around 6 to 8 $M_{\sun}$ are thought to go through a second dredge-up to become asymptotic giant branch stars. We encourage the interested reader to further exploit the potential of our catalog.
}


\maketitle

\section{Introduction}

Massive stars are stars born with initial masses $\gtrsim8~M_{\sun}$. They are relatively rare compared to the large number of low-mass stars due to the initial mass function and their short lifetimes. However, as a result of the intensive interior energy transfer and radiative output, they are responsible for some of the most extreme astrophysics in the Universe, including supernovae (SN), black holes, gravitational waves and long gamma-ray bursts, and are critical for stellar evolution, star formation and chemical evolution throughout the cosmic time \citep{Humphreys1984, Woosley2002, Massey2003a, Meynet2011, Maeder2012, Massey2013}. The first generation of stars in the early Universe is expected to be massive and one of the main contributors of dust content in high-redshift galaxies \citep{Massey2005, Bromm2009, Levesque2010}. Therefore, understanding of the physical properties, evolution and mass loss of massive stars may help to reveal the formation of the primitive cosmic structures \citep{Gall2011, Smith2014, Zhang2018}. 

Unfortunately, such stages of the early Universe can not be directly observed. The alternative way is to study the analogs of early Universe in our own cosmic backyard. In recent years, there has been a growing interest in massive stars in the metal-poor environments, due to the advancement of detectors and telescopes, and the needs to extrapolate the astrophysics from the local Universe to the early cosmic epochs. As a consequence, metal-poor star-forming dwarf irregular (dIrr) galaxies in the local Universe serve as an ideal laboratory for investigating the evolution and mass loss of massive stars at low-metallicity, since they mimic the behaviours of galaxies in the early Universe \citep{Kunth2000, McConnachie2012}.

To understand the evolution of massive stars in the low-metallicity environment, both theoretical and observational constraints are needed. Among the many physical parameters of massive stars, one deterministic parameter is the mass loss, which has a profound impact on a star's lifetime, luminosity (L), effective temperature ($T_{\textrm{eff}}$), radiation field, and the end fate as SN. However, the most important modes of mass loss are the most uncertain \citep{Smith2014}. It has been known that the mass-loss rates (MLR) adopted in modern stellar evolution codes for standard metallicity-dependent winds of hot main-sequence stars are overestimated by a factor of 2-3, due to the clumped and inhomogeneous stellar winds \citep{Puls2008}. Thus, other factors, e.g., the stellar winds, pulsations, rotation, convection and eruptions of evolved supergiants, as well as binary mass transfer, may have important contributions to the removal of the hydrogen envelope. Or, the overestimation of MLR of the hot stars may just simply raise the mass-limit for what stars can become Wolf-Rayet stars (WRs) through stellar wind mass-loss alone. Meanwhile, it has also been recognized that the unsteady modes of mass loss, like the episodic mass-loss events, are more important than previously thought \citep{Smith2006, Ofek2013}. Consequently, observation in the infrared wavelengths, especially the relatively longer wavelengths, which are mainly dominated by dust emission, has a great impact on the understanding of MLR of massive stars.

Among the Local Group galaxies, the Large and Small Magellanic Cloud (LMC and SMC) are particularly intriguing due to their low-metallicity environments (about half and one-fifth of the Milky Way; \citealt{Russell1992, Rolleston2002, Keller2006, Dobbie2014, DOnghia2016}) and close distances, for which individual stars can be resolved, allowing for a detailed analysis of their massive star populations in a variety of ways \citep{Barba1995, Massey2003b, Evans2008, Bonanos2009, Bonanos2010, Neugent2010, Yang2011, Yang2012, Bouret2013, Kourniotis2014, Hainich2015, Castro2018, Yang2018, Britavskiy2019, Patrick2019}. 

In this paper, we focus on the evolved dusty massive star population in the SMC. We construct a clean catalog of massive stars, including their infrared properties (related to the MLR), astrometric solution regarding their membership to the SMC, time-series data revealing stellar variability, and evolutionary models in relation to their evolutionary stages, aiming to build a comprehensive anchor for future studies. The paper is structured as follows: the multiwavelength source catalog and time-series data are presented in \textsection2 and \textsection3, respectively. The identification of evolved massive star candidates is described in \textsection4. The summary is given in \textsection5.  

\section{Multiwavelength Source Catalog}

\begin{figure}
\center
\includegraphics[bb=125 365 455 690, scale=0.65]{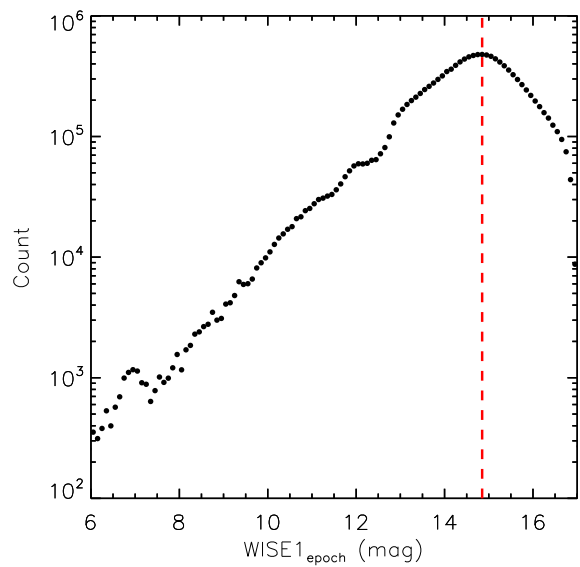}
\caption{Histogram of ALLWISE WISE1 single-epoch measurements in the SMC region. A drop-off of 14.85 mag is shown by the red dashed line. \label{w1_epoch_histo}}
\end{figure}

Since our goal is to focus on the evolved dusty massive stars in the SMC, we use the \textit{Spitzer} Enhanced Imaging Products (SEIP) source list as a starting point. SEIP source list contains sources detected with a high signal-to-noise ratio (S/N; 10-sigma level) in at least one channel among 12 near-infrared (NIR) to mid-infrared (MIR) bands of J (1.25 $\mu$m), H (1.65 $\mu$m), K$_{\rm S}$ (2.17 $\mu$m), IRAC1 (3.6 $\mu$m), IRAC2 (4.5 $\mu$m), IRAC3 (5.8 $\mu$m), IRAC4 (8.0 $\mu$m), MIPS24 (24 $\mu$m), WISE1 (3.4 $\mu$m), WISE2 (4.6 $\mu$m), WISE3 (12 $\mu$m) and WISE4 (22 $\mu$m), from Two Micron All Sky Survey (2MASS; \citealt{Skrutskie2006}), \textit{Spitzer} \citep{Werner2004} and Wide-field Infrared Survey Explorer (\textit{WISE}; \citealt{Wright2010})\footnote{\label{footnote_seip}http://irsa.ipac.caltech.edu/data/SPITZER/Enhanced/SEIP/docs/\\seip\_explanatory\_supplement\_v3.pdf}. We retrieved the initial infrared data from SEIP source list with $3^\circ\leq$ R.A. $\leq25^\circ$, $-75.5^\circ\leq$ Decl. $\leq -70^\circ$ and IRAC1 or WISE1 $\leq$ 15.0 mag, covering almost the whole SMC and also a small part of the Magellanic Bridge (MB). The magnitude cut of IRAC1 or WISE1 $\leq$ 15.0 mag was justified based on a drop-off ($\sim$14.85 mag) in the number counts for 12,748,156 ALLWISE WISE1 single-epoch measurements in the same region as shown in Figure~\ref{w1_epoch_histo}. Considering the lower angular resolution of \textit{WISE} ($\sim6''$) and 2MASS ($\sim5''$) compared to \textit{Spitzer} ($\sim2''$), and \textit{WISE} sources within $3''$ of a SEIP source were reported, a deblending was performed with a search radius of $3''$. Targets having neighbors within $3''$ were excluded from the initial dataset, which resulted in 131,233 targets. 

Besides the infrared detection, the reliable membership to the SMC is also a crucial factor for our study, for which the astrometric solution from \textit{Gaia} Data Release 2 (DR2) is vital \citep{Gaia2016, Gaia2018a}. Given the very small offset between the \textit{Gaia} DR2 and the 2MASS (median value of $\sim0.120''\pm0.157''$ for targets with $G\leq18~mag$ in the same SMC region), as well as the SEIP source list and the 2MASS (median value of $\sim0.089''\pm0.237''$\footref{footnote_seip}), we crossmatched the deblended SEIP data and \textit{Gaia} DR2 with a search radius of $1''$ to fix the position, and removed any SEIP targets with multiple counterparts to eliminate the blending. Then, a $3''$ crossmatching was performed and SEIP targets with multiple counterparts were removed again, which resulted in 74,237 targets. Since the effective angular resolution of the \textit{Gaia} DR2 source list has improved to $\sim0.4''$, with incompleteness in close pairs of stars starting below about $2''$ \citep{Gaia2018a, Arenou2018}, we may reasonably believe that the vast majority of the blending is removed at the resolutions of both \textit{Spitzer} and \textit{Gaia}. 

\begin{figure*}
\includegraphics[bb=115 375 485 695, scale=0.47]{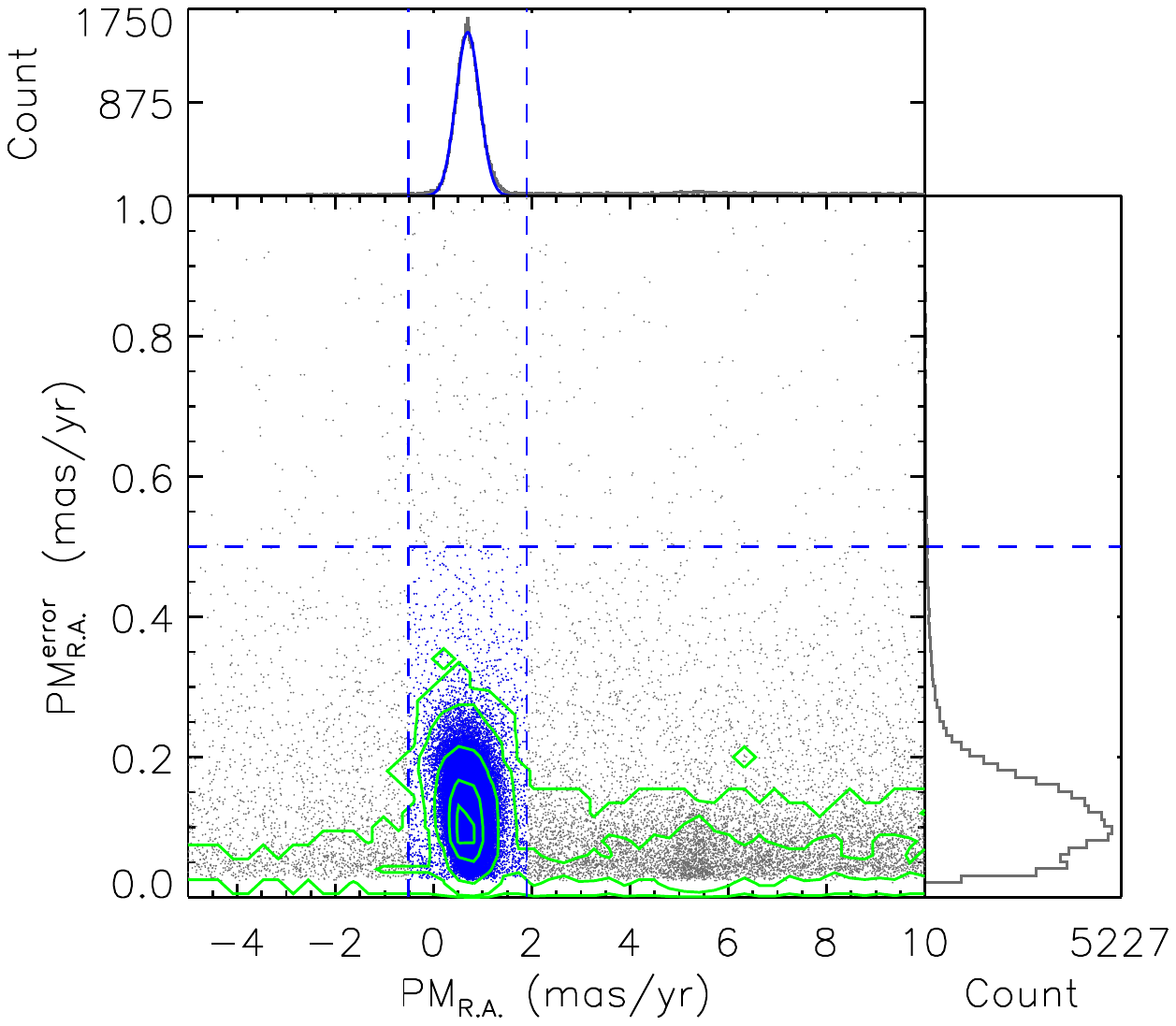}
\includegraphics[bb=115 375 485 695, scale=0.47]{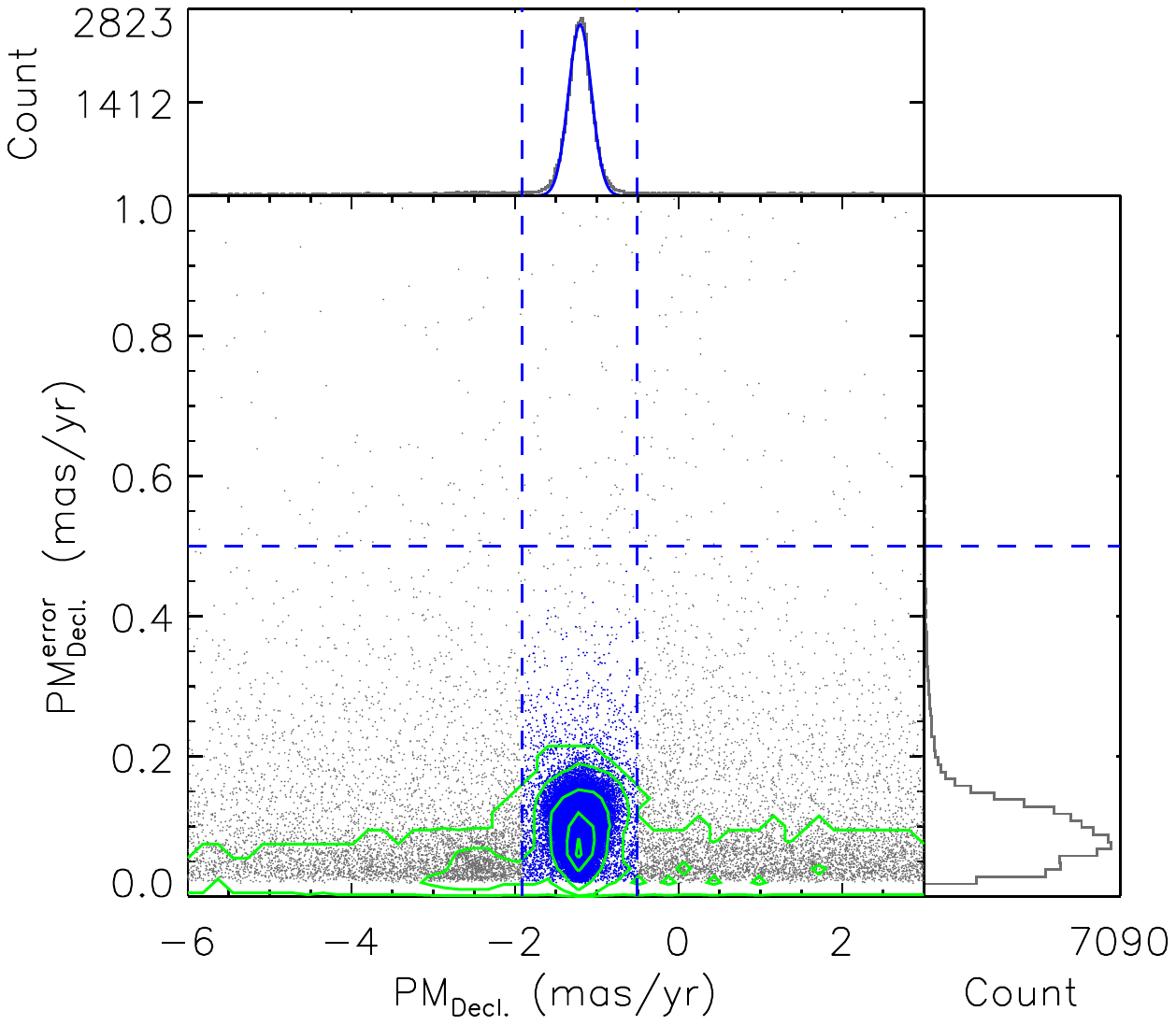}
\includegraphics[bb=120 375 485 695, scale=0.47]{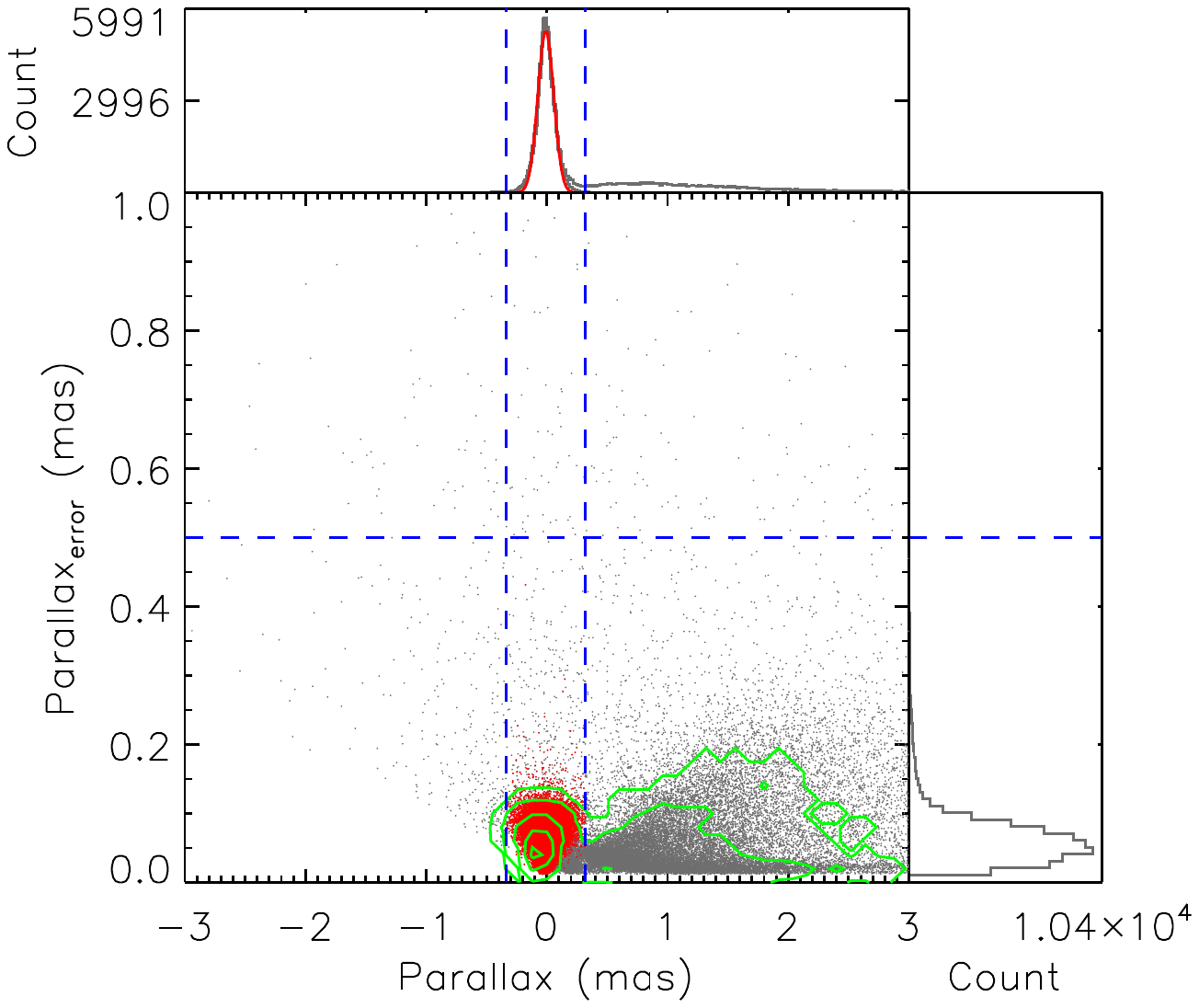}
\caption{Evaluation of the \textit{Gaia} astrometric solution. The limits of PM error (0.5 mas/yr) and parallax error (0.5 mas) are shown as horizontal dashed lines. The first two panels show errors versus \textit{Gaia} PMs in R.A. (left) and Decl. (middle), respectively. A Gaussian profile is fitted to PM in each dimension and the limits of $\pm5\sigma$ are calculated (vertical dashed lines). The selected targets are shown in blue color. The last panel (right) shows the errors versus \textit{Gaia} parallaxes. A Gaussian fitting is adopted again for the parallax, while an additional elliptical constraint is also applied with the $5\sigma$ limits of $PM_{R.A.}$ and $PM_{Decl.}$ taken as the primary and secondary radii, respectively. The same criteria of $\pm5\sigma$ are calculated for the parallax (vertical dashed lines). The final selected targets are shown in red color. Green contours show the number density in each digram. \label{pm_constrain}}
\end{figure*}

After crossmatching and deblending between SEIP and \textit{Gaia} data, the \textit{Gaia} DR2 astrometric solution was used to eliminate the foreground contamination \citep{Lindegren2018}. The selection of SMC members was restricted to targets with errors $<$0.5 mas/yr in proper motion (PM) and $<$0.5 mas in parallax. The first two panels of Figure~\ref{pm_constrain} show the errors versus \textit{Gaia} PMs in R.A. (left) and Decl. (middle), respectively. We fit a Gaussian profile to PM at each dimension as $0.695~(peak)\pm0.240~(\sigma)~mas/yr$ in R.A. and $-1.206\pm0.140~mas/yr$ in Decl., and calculated the limits of $\pm5\sigma$ shown as the vertical dashed lines. The last panel (right) of Figure~\ref{pm_constrain} shows the errors versus \textit{Gaia} parallaxes. Similarly, a Gaussian profile fitting was adopted for the parallax as $-0.009\pm0.066~mas$, while an additional elliptical constraint was also applied with the $5\sigma$ limits of $PM_{R.A.}$ and $PM_{Decl.}$ taken as the primary and secondary radii, respectively. The same criteria of $\pm5\sigma$ were calculated for the parallax. In brief, the membership of the SMC is constrained by a Gaussian profile of the parallax with an additional elliptical constraint derived from $PM_{R.A.}$ and $PM_{Decl.}$, which results in 45,466 targets. Additionally to the previous astrometric constraints, we also applied another constraint on the radial velocity (RV) for all targets that have \textit{Gaia} RV measurements, set at RV being larger than $\sim$90 km/s as shown below.

Further evaluation of the astrometric excess noise, which measures the disagreement, expressed as an angle, between the observations of a source and the best-fitting standard astrometric model (using five astrometric parameters), shows that about 99.5\% of our targets have $astrometric\_excess\_noise\leq0.5$ mas, indicating good astrometric solutions for the vast majority of our targets \citep{Lindegren2018}. Moreover, \citet{Gaia2018b} determined the PMs and and parallaxes for 75 Galactic globular clusters, nine dwarf spheroidal galaxies, one ultra-faint system, and the MCs, where the method was also used by others (see, e.g., \citealt{Aadland2018}). The basic idea was to first determine the median and robust scatter in PMs and parallaxes by selecting a sample of stars covering a larger field of view (FoV), then further eliminated any sources showing larger scatter in the PMs, and finally constructed a filter based on a covariance matrix of the cleaned sample, which allows one to properly select out likely members. They also provided lists of possible members according to their analysis for all the systems. We crossmatched between our list of 45,466 targets and their list of $\sim$1.4 million sources for the SMC (their data covered larger FoV, which also yielded more identified sources than ours) with a search radius of 1''. The result was extremely good. Out of 45,466 targets in our list, there was only one mismatched target. We also compared their result with our LMC source catalog (in preparation), where 99.98\% of our targets were matched. This proved the robustness of our method in selection of extragalactic objects by using \textit{Gaia} astrometric solution, even without any sophisticated correction or filtering (in general, our method was actually similar to theirs). There were likely four reasons for this. Firstly, we put relatively strict constraints on the \textit{Gaia} astrometric solution. Secondly, the SMC was close enough to have better quality of the data compared to more distant galaxies. Thirdly, we presumed that, for the current stage, compared to the Galactic targets, extragalactic objects (even as close as in the MCs) were probably only seen as tiny point sources by \textit{Gaia}, where the fluctuation of photocenter caused by the binarity or convection/pulsation were almost invisible \citep{Chiavassa2011, Pasquato2011, Messineo2019}. Fourthly, the crowding issue was largely mitigated since we deblended our sample. However, we would also like to emphasize that, in some rare cases, \textit{Gaia} PMs and parallaxes might not good enough to distinguish a halo giant (e.g., at a distance of 10-15 kpc) from an SMC supergiant (at a distance of $\sim$60 kpc). Meanwhile, the RV might be not very helpful either, since the SMC's RV was mostly a reflection of the Sun's motion around the the Milky Way, and hence a halo star in the same direction might have a similar RV as an SMC member. Even with spectroscopy, it would be hard to distinguish, as a metal-poor giant in the halo might have similar metallicity as an SMC member. Therefore, we used the Besan\c{c}on models of the Milky Way \citep{Robin2003} to generate the expected number of Galactic stars as a function of magnitude and color for the same area of the sky as the SMC. The result of simulation indicated that, after applying the same constraints as above, the contamination by the halo stars was less than 0.25\% in our sample and could be ignored.

\begin{figure}
\includegraphics[bb=125 365 465 690, scale=0.7]{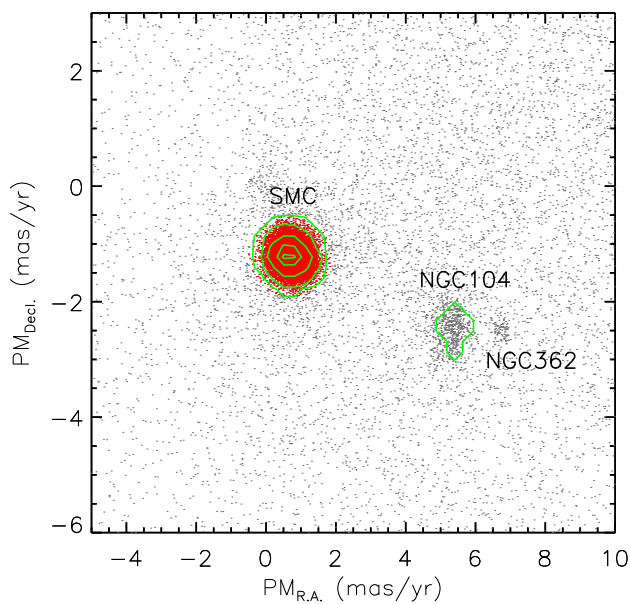}
\caption{$PM_{R.A.}$ versus $PM_{Decl.}$ diagram, in which the separation of selected SMC members (red), NGC104 and NGC362 is clearly shown. Based on this diagram, we estimate the contamination of remaining foreground sources for the SMC is around 0.2\% ($\sim$98/45,466) and can be ignored. Green contours represent the number density. \label{pm_radec}}
\end{figure}

Figure~\ref{pm_radec} shows $PM_{R.A.}$ versus $PM_{Decl.}$, for which the separation of selected SMC members, NGC104 and NGC362 is clearly shown. Meanwhile, based on the median number density of Figure~\ref{pm_radec} and the association between the SEIP source list and the 2MASS point source catalog (SEIP sources without valid 2MASS measurements), we estimated the contamination of remaining foreground sources and the possible non-point/background sources for the SMC were around 0.2\% ($\sim$98/45,466) and 0.3\% ($\sim$131/45,466), respectively, and could be ignored. Background point-like sources, for example, active galactic nuclei, quasars, or blue compact dwarfs, cannot be rejected at this stage, while the non-detection in 2MASS may also be due to the saturation, faintness, or photometric qualities of the targets. The estimation of contamination indicates that about 99.5\% of the targets in our source catalog are most likely genuine members of the SMC. 

\begin{figure}
\includegraphics[bb=100 430 460 615, scale=0.7]{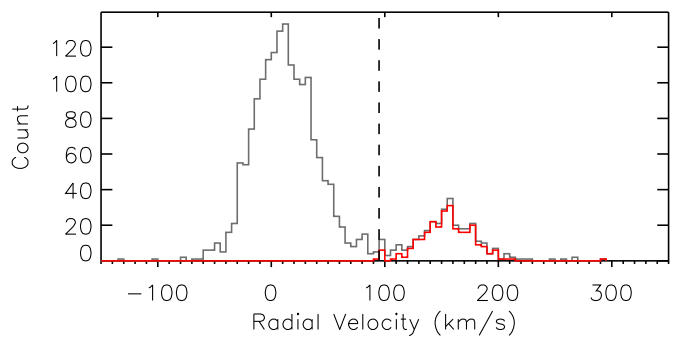}
\caption{Histogram of RVs from \textit{Gaia}. The separation of the Milky Way and the SMC is clear, and the vast majority of targets with RVs larger than $\sim$90 km/s are selected (red) with minimal value of $\sim$95 km/s (dashed line). \label{gaia_rv}}
\end{figure}

\begin{figure}
\includegraphics*[bb=125 365 495 690, scale=0.67]{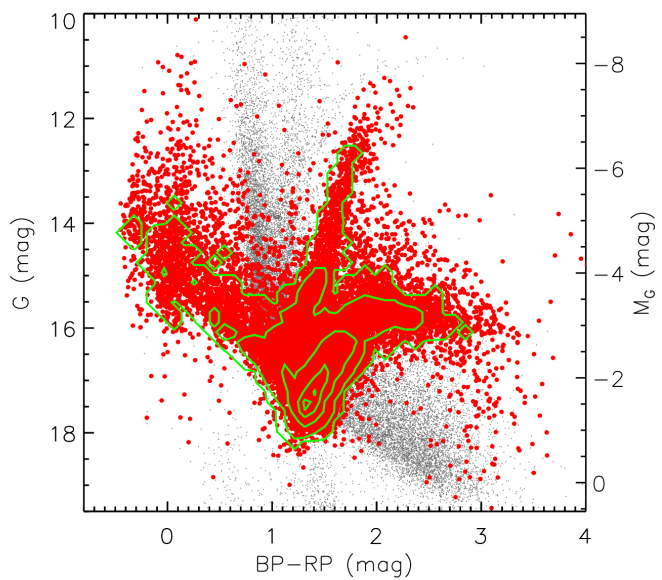}
\caption{G versus BP-RP diagram for the \textit{Gaia} data before (gray) and after (red) the astrometric constraints, where the large number of foreground contamination is swept out. Green contours represent the number density. \label{gaia_cmd}}
\end{figure}

We note that the strict constraints on the astrometric solution and the previous deblending procedure may cause target loss and incompleteness in our catalog to a certain extent, but ensure that we select the true SMC targets. This also can be seen from the histogram of \textit{Gaia} RVs in Figure~\ref{gaia_rv}, where the separation of Milky Way and SMC is clear, and the vast majority of astrometry constrained targets with \textit{Gaia} RVs larger than $\sim$90 km/s are selected with minimal value of $\sim$95 km/s (targets with both large PMs and RVs are not necessarily members of the SMC, as they could be hypervelocity stars in the Milky Way, runaway stars from the Milky Way or the SMC, or free-floating stars between the Milky Way and the SMC). Figure~\ref{gaia_cmd} illustrates the \textit{Gaia} color-magnitude diagram (CMD) before (gray) and after (red) applying the astrometric constraints, where the large amount of foreground contamination of bright yellow and faint red stars is swept out. In addition, further constraints on the SEIP data, for example, $S/N\geq3$ for IRAC1 ($i1\_fluxtype=1$), were investigated and resulted only in about 1.66\% difference in the total number of sources (45,466 versus 44,712 targets) and were ignored.

\begin{figure*}
\center
\includegraphics[bb=55 365 560 710]{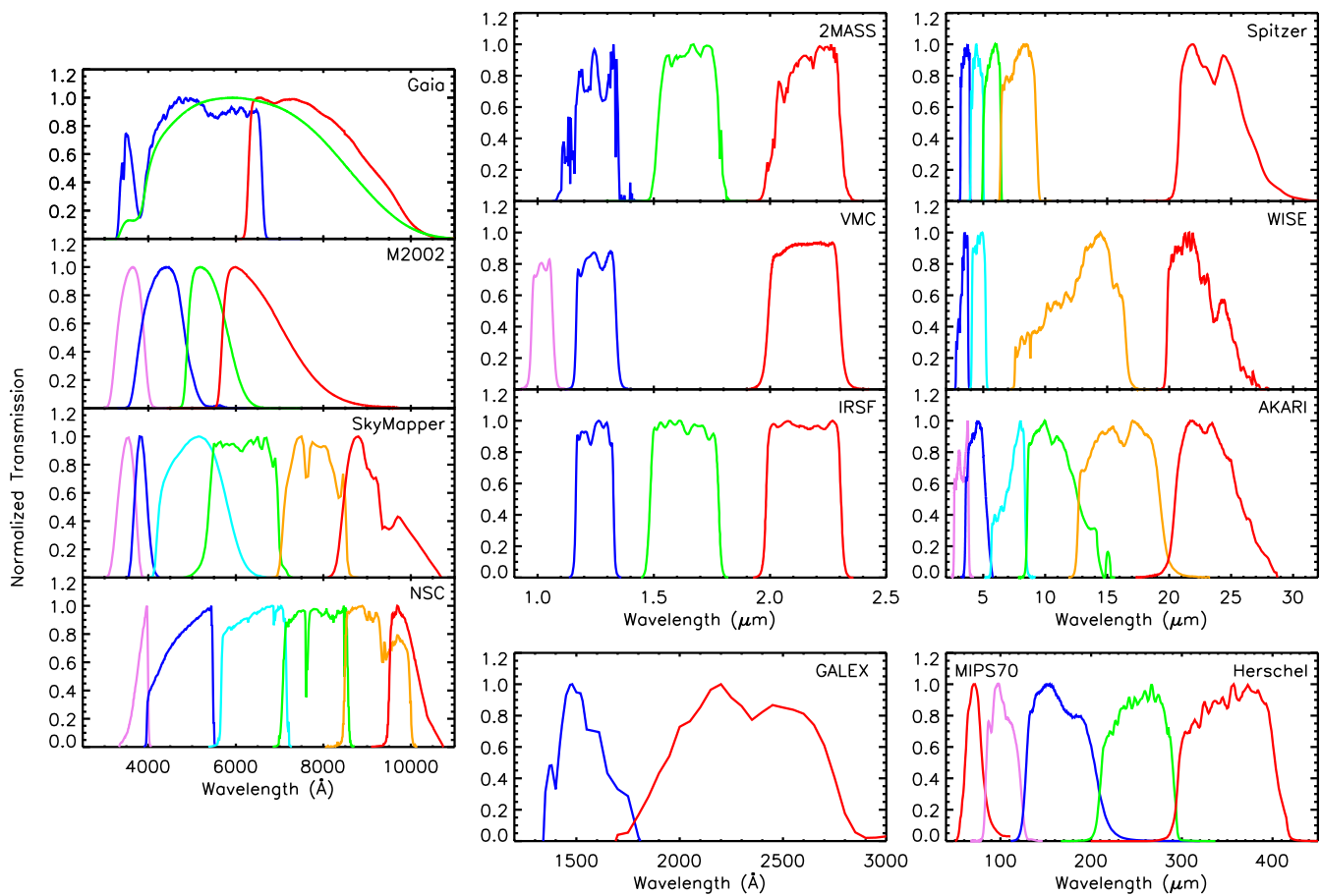}
\caption{Normalized transmission curves of filters (convolved with the instrument/telescope sensitivities) used in our study. In total, we have 50 filters including 21 optical (includes the two UV filters) and 29 infrared filters. For each dataset, the filters are color coded from shorter (blueish) to longer (reddish) wavelengths. Left panel from top to bottom: BP, RP, G from \textit{Gaia}, U, B, V, R from M2002, $u, v, g, r, i, z$ from SkyMapper, and $u, g, r, i, z$, Y from NSC. Middle panel from top to bottom: J, H, K$_{\rm S}$ from 2MASS, Y, J, K$_{\rm S}$ from VMC, J, H, K$_{\rm S}$ from IRSF, and FUV, NUV from \textit{GALEX}. Right panel from top to bottom: IRAC1, IRAC2, IRAC3, IRAC4, MIPS24 from \textit{Spitzer}, WISE1, WISE2, WISE3, WISE4 from \textit{WISE}, N3, N4, S7, S11, L15, L24 from \textit{AKARI}, and MIPS70 (from \textit{Spitzer}), PACS100, PACS160, SPIRE250, SPIRE350 from \textit{Herschel}. \label{filters}}
\end{figure*}

\begin{figure*}
\center
\includegraphics*[bb=50 395 540 660, scale=0.9]{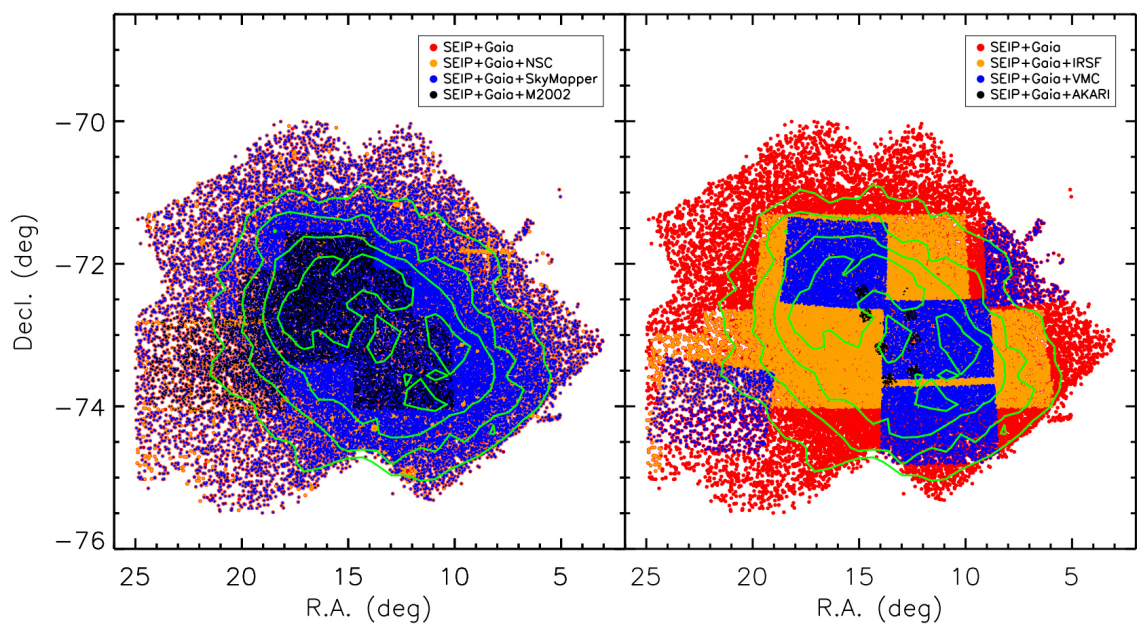}
\caption{Spatial distribution of the additional optical (left) and infrared (right) datasets. The \textit{GALEX} and HERITAGE data are not shown in the diagram due to the paucity of matches. Green contours represent the number density. \label{spatial}}
\end{figure*}

Based on this fiducial dataset of 45,466 targets, we retrieved additional optical and infrared data from the following datasets with a search radius of $1''$:
\begin{itemize}[noitemsep,topsep=0pt,parsep=0pt,partopsep=0pt] 
	\item 18,641 matches (41.00\%) from VISTA survey of the Magellanic Clouds system (VMC) DR4. VMC is a NIR YJK$_{\rm S}$ bands multi-epoch survey for the LMC, the SMC and the MB (K$_{\rm S}$ $\lesssim$ 20.3 mag, $\sim$2\% photometric and $\sim0.01''$ astrometric precision), using the 4-meter near-infrared optimized Visible and Infrared Survey Telescope for Astronomy (VISTA; \citealt{Cioni2011}).
	\item 28,678 matches (63.08\%) from the IRSF Magellanic Clouds point source catalog (MCPS). IRSF MCPS is a NIR JHK$_{\rm S}$ bands photometric catalog for a 40 $deg^2$ area in the LMC, an 11 $deg^2$ area in the SMC, and a 4 $deg^2$ area in the MB (K$_{\rm S}$ $\lesssim$ 16.6 mag, photometric and astrometric accuracies for bright sources are 0.03-0.04 mag and $0.1''$, respectively), based on the data from Simultaneous three-color InfraRed Imager for Unbiased Survey (SIRIUS) camera on the InfraRed Survey Facility (IRSF) 1.4-meter telescope \citep{Kato2007}.
	\item 625 matches (1.37\%) from \textit{AKARI} SMC bright point source list. The source list represents NIR to MIR imaging and spectroscopic observations of patchy areas in the SMC (N3 $\lesssim$ 16.5 mag, $\sim0.1~mag$ photometric and $\lesssim0.8''$ astrometric precision), using the Infrared Camera (IRC) aboard \textit{AKARI} space telescope \citep{Onaka2007, Murakami2007, Ita2010}.
	\item 4 matches from HERschel Inventory of the Agents of Galaxy Evolution (HERITAGE) band-merged  source catalog (units are in flux [mJy] instead of magnitude). HERITAGE catalog is derived based on the data from both Photodetector Array Camera and Spectrometer (PACS; 100 and 160$\mu$m) and Spectral and Photometric Imaging Receiver (SPIRE; 250, 350, and 500$\mu$m) cameras on board the \textit{Herschel} Space Observatory, to identify dusty objects in the LMC and SMC (the catalog also includes the \textit{Spitzer} MIPS 70$\mu$m band data, and none of our targets has been detected in the SPIRE 500$\mu$m band; \citealt{Pilbratt2010, Meixner2013, Seale2014}).
	\item 40,387 matches (88.83\%) from SkyMapper DR1.1 (we apply constraints on the parameters as $flags<4$, $nch\_max=1$, $nimaflags=0$ and $class\_star\geq0.9$ to retrieve the data; see \citealt{Wolf2018} for details). SkyMapper is a southern hemisphere photometric survey in six bands ($u, v, g, r, i, z$; from magnitude 8 to 18, $\sim$1\% photometric and $<0.2''$ astrometric precision), using the dedicated 1.3-meter SkyMapper telescope \citep{Keller2007, Bessell2011, Wolf2018}. 
	\item 38,759 matches (85.25\%) from the NOAO source catalog (NSC) DR1 (we apply constraints on the parameters as $flags<4$ and $class\_star\geq0.9$ to retrieve the data; see \citealt{Nidever2018} for details). NSC is a catalog of sources from most of the public data taken on NOAO's CTIO-4m+DECam as well as KPNO-4m+Mosaic3 (in $u, g, r, i, z$, Y bands; reach to $\sim$23 mag in most broadband filters with $\sim$1-2\% photometric precision, and astrometric accuracy of $\sim$7 mas; \citealt{Nidever2018}). 
	\item 11,630 matches (25.58\%) from a UBVR CCD survey of the Magellanic Clouds by \citet{Massey2002} (M2002; note that due to the existence of some extreme large errors in the catalog, we replace them, e.g., error $>$ 1.0 mag, with NULL values). M2002 is a survey of 14.5 $deg^2$ region in the LMC and 7.2 $deg^2$ region in the SMC (V $\lesssim$ 18.0 mag, $<3\%$ photometric and $\sim0.3''$ astrometric precision), using 0.61-meter Curtis Schmidt telescope at CTIO. 
	\item 164 matches (0.36\%) from revised \textit{GALEX} source catalog for the All-Sky Imaging Survey (GUVcat\_AIS). GUVcat\_AIS is a science-enhanced, ``clean'' catalog of \textit{GALEX} ultraviolet (UV) sources with typical depth of 20.8 and 19.9 mag, position accuracy of $0.32''$ and $0.34''$ in NUV and FUV bands, respectively \citep{Morrissey2007, Bianchi2017}.
\end{itemize} 
All the datasets are also deblended with a search radius of $3''$, prior to crossmatching with the SEIP-Gaia dataset. Figure~\ref{filters} shows the normalized transmission curves of all filters used in our study. In total, we have 50 filters including 21 optical (including the two UV filters) and 29 infrared filters. The spatial distributions of the additional optical (left) and infrared (right) datasets are shown in Figure~\ref{spatial}. The \textit{GALEX} and HERITAGE data are not shown in the diagram due to the paucity of matches.

Besides the photometric data, additional classifications were also retrieved from the literature with a search radius of $1''$, including:
\begin{itemize}[noitemsep,topsep=0pt,parsep=0pt,partopsep=0pt] 
	\item 37,375 matches from \citet{Boyer2011} (IR color classifications). It was an investigation of the IR properties of cool, evolved stars in the SMC, including the red giant branch (RGB) stars and the dust-producing red supergiant (RSG) and asymptotic giant branch (AGB) stars using observations from the \textit{Spitzer} Space Telescope Legacy Program entitled ``Surveying the Agents of Galaxy Evolution in the Tidally Stripped, Low Metallicity SMC (SAGE-SMC)''.
	\item 30 matches from \citet{Sewilo2013} (IR color and SED classifications). They used CMDs based on the multiwavelength photometric data and SED fitting to identify a population of $\sim$1,000 intermediate- to high-mass young stellar objects (YSOs) in the SMC.
	\item 43 matches from \citet{Ruffle2015} (MIR spectral classifications). They classified 209 point sources observed by \textit{Spitzer} Infrared Spectrograph (IRS; \citealt{Houck2004}) using a decision tree method, based on infrared spectral features, continuum and spectral energy distribution shape, bolometric luminosity, cluster membership and variability statistics (all the targets from \citealt{Kraemer2017} were also included). 
	\item 695 matches from \citet{Bonanos2010} (optical spectral classifications). It was a catalog of 5,324 massive stars from the literature with accurate spectral types, and a multiwavelength photometric catalog for a subset of 3,654 of these stars in the SMC, intending to study their infrared properties.
	\item 198 matches from \citet{Gonzalez2015} (spectral variability flag, radial velocities and optical spectral classifications; for convenience, we only kept the first spectral classification for targets with multiple measurements). They studied physical properties of about 500 RSGs in the LMC and SMC by using NIR/MIR photometry and optical spectroscopy, aiming at exploring the fainter end of RSGs and extrapolating their behavior to other environments by building a more representative sample. 
	\item 113 matches from \citet{Neugent2018} (data originated from \citealt{Neugent2010}; radial velocities and optical spectral classifications). They spectroscopically observed 176 near-certain (Category 1) SMC yellow supergiant stars (YSGs) among $\sim$500 candidates to test against the evolutionary model in the low-metallicity environment.
	\item 39,295 matches from Simbad \citep{Wenger2000}. The radial velocities, optical spectral classifications, main object types, and auxiliary object types were retrieved.
\end{itemize}
The unmatched targets are likely due to larger PMs, blends, or the quality cuts in SEIP or \textit{Gaia} catalog.

\begin{figure*}
\center
\includegraphics[bb=55 365 560 710]{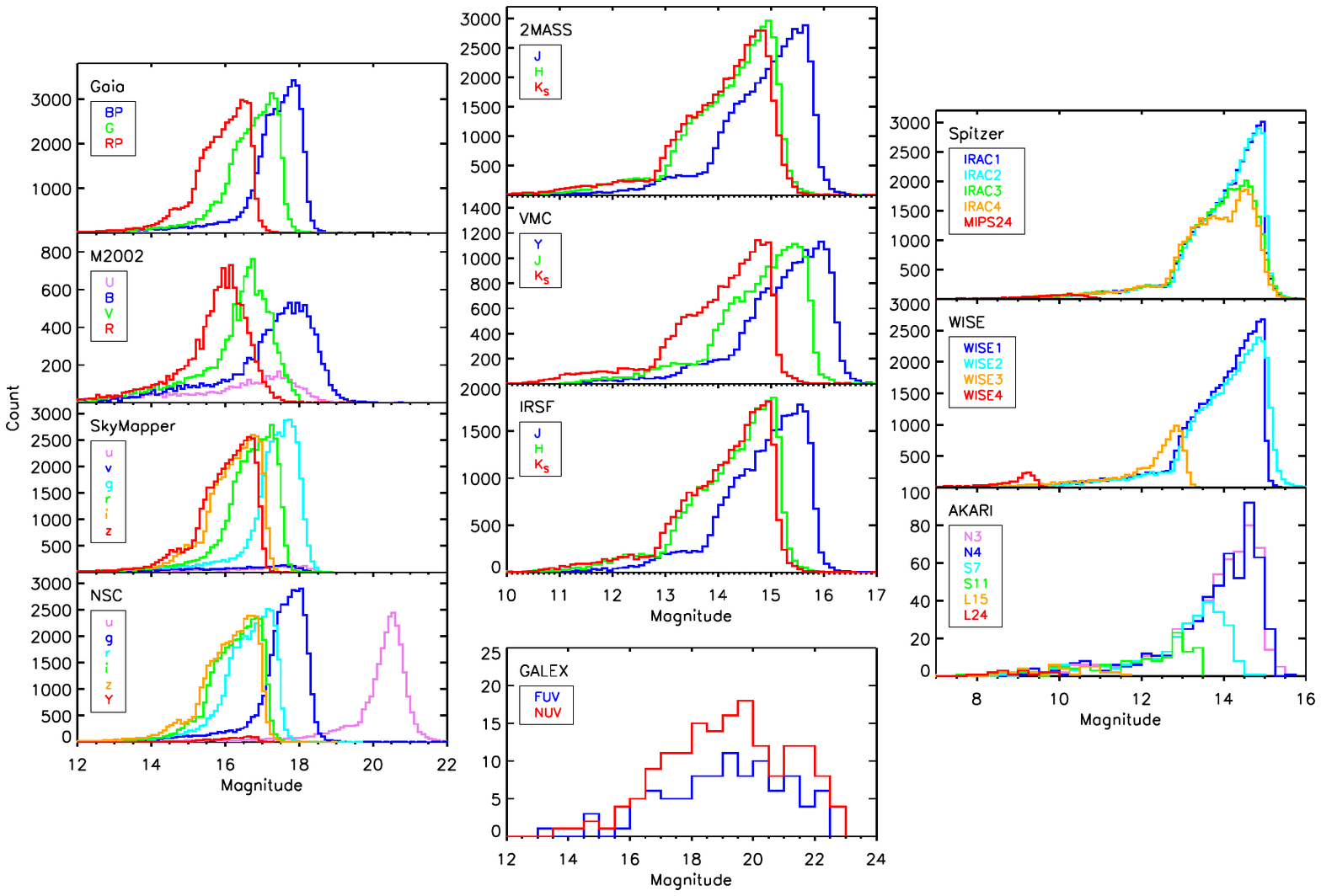}
\caption{Histograms of magnitude distribution in each dataset. The bin size of magnitude is 0.1 mag, except for the \textit{GALEX} (0.5 mag) and \textit{AKARI} (0.25 mag) data. For WISE3 and WISE4 bands, the histograms only show targets with $S/N\geq2$. For convenience, the HERITAGE data are not shown in the diagram. \label{sc_histo}}
\end{figure*}

This multiwavelength source catalog with 45,466 targets serves as the backbone of our study. The sample consists of ``bona-fide'' and ``dusty'' SMC targets determined by both astrometric measurement and infrared detection. Table~\ref{fpercent} shows the absolute and relative percentage (relative to the filter with the most matches in each dataset) of detected targets in each filter. Figure~\ref{sc_histo} shows the histograms of magnitude distribution for each dataset (for convenience, the HERITAGE data is not shown here). The bin size is 0.1 mag, except for the \textit{GALEX} (0.5 mag) and \textit{AKARI} (0.25 mag) data. For WISE3 and WISE4 bands, due to the fact that the majority of the targets ($\sim$75\% in WISE3 and $\sim$95\% in WISE4 bands) have low S/N ($<2$) and are derived with a 95\% confidence brightness upper limit, the histograms only show targets with $S/N\geq2$. 

\section{Multiwavelength Time-Series Data}

\begin{figure*}
\center
\includegraphics*[bb=60 545 555 720]{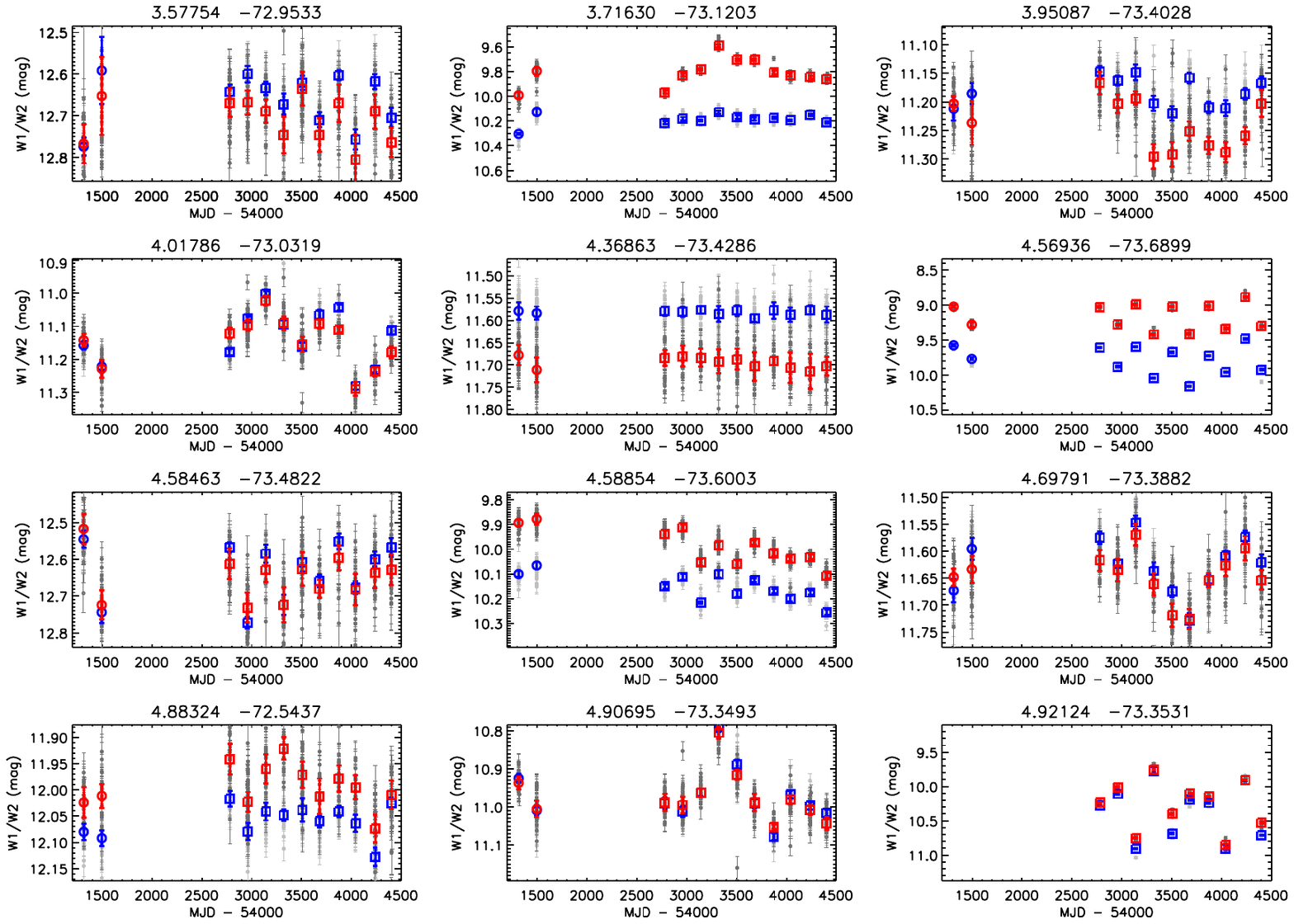}
\caption{Examples of original (light gray for WISE1 and dark gray for WISE2) and binned (blue for WISE1 and red for WISE2) lightcurves. There are about twelve major epochs with two epochs from ALLWISE (open circles) and ten epochs from NEOWISE-R (open squares) separated by an approximately three-year gap. Coordinates of the targets are indicated on top of each panel. \label{lc_group}}
\end{figure*}

Following \citet{Yang2018}, the MIR time-series data of WISE1 (3.4 $\mu$m) and WISE2 (4.6 $\mu$m) bands for all the 45,466 targets were collected from both ALLWISE \citep{Cutri2013} and Near-Earth Object \textit{WISE} Reactivation mission (NEOWISE-R; \citealt{Mainzer2014}), with a search radius of $1''$ and the following parameter constraints, $qi\_fact>0$, $saa\_sep>0$, $moon\_masked=0$, $qual\_frame>0$, $det\_bit=3$, $\chi^2\leq10$, $S/N\geq3$ (see \citealt{Yang2018} for details), which resulted in about 2.28 million measurements from ALLWISE and 8.78 million measurements from NEOWISE-R. With the NEOWISE 2019 data release, the total frame coverage is about twelve major epochs spanning $\sim$3,200 days ($\sim$8.8 years) with two epochs from ALLWISE and ten epochs from NEOWISE-R separated by an approximately three-year gap. The beginning and ending of each epoch set by us are given in Table~\ref{epochtbl}. We binned the data within each epoch by using the median values of the date and magnitudes. For each epoch, we required at least five valid points to calculate the median value. Figure~\ref{lc_group} shows examples of the original lightcurves overlapped with the binned lightcurves. The median absolute deviation (MAD) and standard deviation (SD) were used to calculate the long-term (full lightcurve) and short-term (within single epoch) variability of each target with at least five valid points, where the former was more resistant to outliers than the latter \citep{Rousseeuw1993}. Although it is possible to derive periods based on the current data, the period search may be highly contaminated by the strong alias structures due to the very low sampling of \textit{WISE} data as shown in Figure~\ref{lc_group}, that we only have twelve epochs spanning $\sim$3200 days and each epoch only covers about 5 to 10 days (we refer interested readers to \citealt{Chen2018} for the \textit{WISE} catalog of periodic variable stars). In addition to MAD and SD, the full amplitude ($\rm Amp=max_{mag}-min_{mag}$) was also calculated for each target. More details about the \textit{WISE} time-series data reduction can be found in \citet{Yang2018}. In total, there are 39,694 (87.30\%) targets having variability statistics in both WISE1 and WISE2 bands covering almost the whole area of our sample. The lack of variability statistics for some targets is likely due to either faint magnitudes or the quality cuts we adopted.

\begin{figure*}
\includegraphics*[bb=65 445 550 620, scale=1.05]{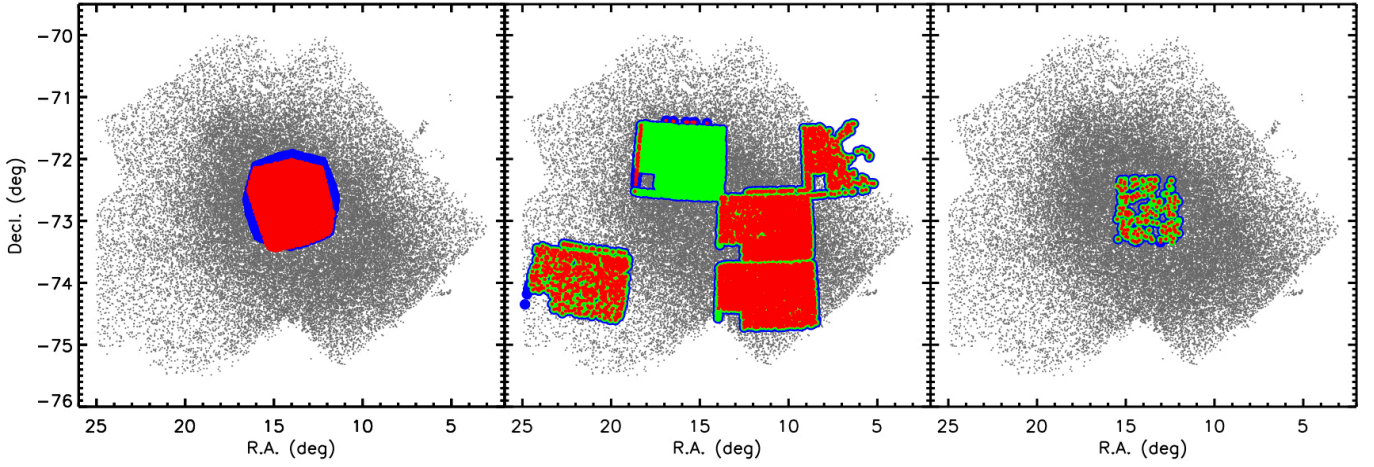}
\caption{Spatial distribution of targets with IR variability statistics matched with our source catalog within $1''$. Left: targets from SAGE-Var project, where blue and red colors indicate the targets in IRAC1 and IRAC2 bands, respectively. Middle: targets from VMC DR4, where blue, green and red colors indicate the targets in Y, J, and K$_{\rm S}$ bands, respectively. Right: targets from IRSF survey, where blue, green and red colors indicate the targets in J, H, and K$_{\rm S}$ bands, respectively. \label{ir_var_plot}}
\end{figure*}

Besides the \textit{WISE} data, we also collected other sets of variability data from different projects, including IR data from SAGE-Var program \citep{Riebel2015}, VMC, IRSF, and optical data from \textit{Gaia}, NSC and OGLE. SAGE-Var is a follow-up to the \textit{Spitzer} legacy program SAGE \citep{Meixner2006, Gordon2011}, for which six total epochs of photometric observations at IRAC1 (3.6 $\mu$m) and IRAC2 (4.5 $\mu$m) bands are obtained covering the bar of the LMC and the central region of the SMC with 15 different timescales ranging from $\sim$20 days to $\sim$5 yr. We collected the SMC data and calculated the median magnitudes, MADs, SDs and Amps for targets with all six epochs available. In total, there are 7,160 (15.75\%) targets in IRAC1 band and 5,894 (12.96\%) targets in IRAC2 band matched with our source catalog within $1''$. The left panel of Figure~\ref{ir_var_plot} shows the spatial distribution of those targets overlapped on our source catalog.
 
We currently rely on the available information from VMC DR4 regarding the variability statistics, as the final release (which may also include the time-series data) with global photometric and astrometric calibration will be made upon completion of the survey\footnote{http://horus.roe.ac.uk/vsa/}. Since the observational cadences for different targets and filters were irregular and the median values of cadences varied from few hours to hundreds of days, for each filter, we constrained all the values of median magnitudes, MADs, SDs, and Amps to be within the range of 0 to 99 with at least five good measurements ($*nGoodObs\geq5$) to avoid any unphysical values. In total, there are 11,197 (24.63\%) targets matched with our source catalog within $1''$ as shown in the middle panel of Figure~\ref{ir_var_plot}. There are 11,197 (24.63\%) targets in Y band, 11,125 (24.47\%) targets in J band, and 7,574 (16.66\%) targets in $K_S$ band. Additionally, there are 204 targets that have been classified as Cepheids by the VMC team matched with our source catalog within $1''$.

For IRSF data, we retrieved the time-series data of $\sim$1,000 targets for each filter from \citet{Ita2018}, where a very long-term (2000-2017) near-infrared variable star survey towards an area of $3~deg^2$ along the bar in the LMC and an area of $1~deg^2$ in the central part of the SMC was carried out with more than one hundred times repeated observations for each area. The median magnitudes, MADs, SDs, Amps were calculated for each targets. There are 160 targets in J band, 161 targets in H band, and 154 targets in K$_{\rm S}$ band matched with our source catalog within $1''$ as shown in the right panel of Figure~\ref{ir_var_plot}.

\begin{figure*}
\includegraphics*[bb=65 445 550 620, scale=1.05]{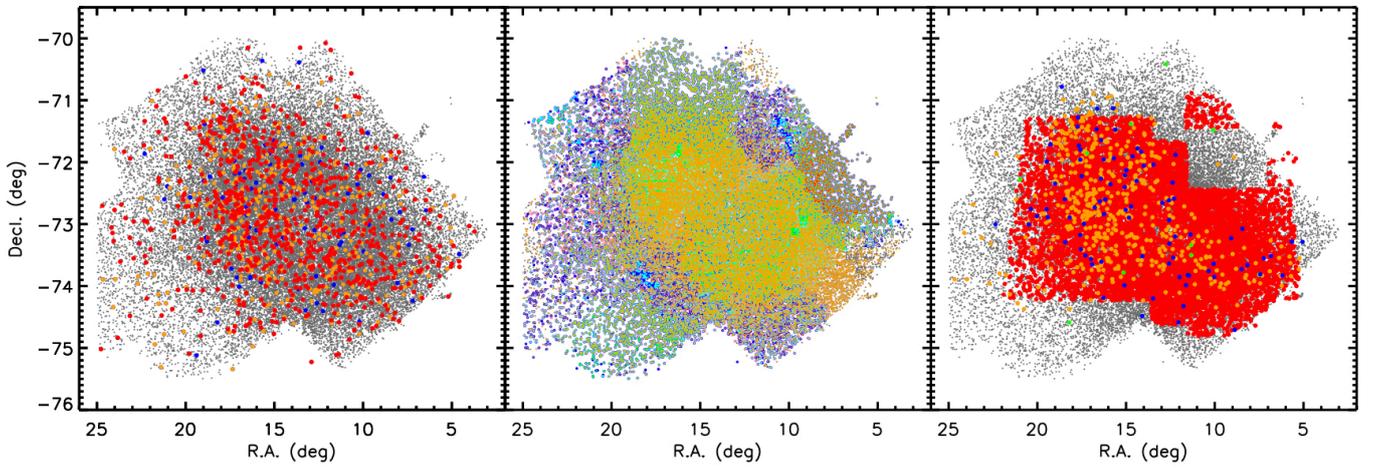}
\caption{Spatial distribution of targets with optical variability statistics matched with our source catalog within $1''$. Left: targets from Gaia, where blue, orange and red colors indicate targets without classification, with $best\_class\_score<0.5$, and $best\_class\_score\geq0.5$, respectively. Middle: targets from NSC, where violet, blue, cyan, green, orange and red colors indicate the targets in $u, g, r, i, z$, and Y bands, respectively. Right: targets from OGLE, where blue, green, orange, and red colors indicate the targets classified as Ecls, T2Ceps, CCeps, and LPVs, respectively. \label{optical_var_plot}}
\end{figure*}

\textit{Gaia} DR2 provides classifications for more than 550,000 variable sources consisting of different types of variables. However, only a subset of the variable stars classified as a certain type are characterized in detail and a fraction of the classifications may well be wrong \citep{Eyer2019, Mowlavi2018}. Since the time-series data would be provided in the future release, we retrieved the variability statistics (including MADs, SDs and Amps) and classifications from \textit{Gaia} Archive\footnote{https://gea.esac.esa.int/archive/} with typically $\sim$30 measurements spanning $\sim$620 days. However, we note that the MAD is scaled by 1.4826, so that the expectation of the scaled MAD at large number of measurements equals the standard deviation of a normal distribution\footnote{https://gea.esac.esa.int/archive/documentation/GDR1/datamodel\\/Ch2/phot\_variable\_time\_series\_gfov\_statistical\_parameters.html}. For consistency of the dataset, we reversed the scaled MAD by dividing it with 1.4826. There are 1,379 (3.03\%; blue) targets with variability statistics which are matched with our source catalog within $1''$ as shown in the left panel of Figure~\ref{optical_var_plot}, while 1,277 (2.81\%; orange) of them are classified, and 868 (1.91\%; red) of them having $best\_class\_score\geq0.5$ ($best\_class\_score$ is a quantity between 0 and 1 provided by the classification pipeline to estimate the confidence of the classifier in the identification of various variability types), for which the classification may be acceptable.

The NSC time-series data of the SMC are mainly from the Survey of the Magellanic Stellar History (SMASH; \citealt{Nidever2017}). We retrieved the data from NOAO Data Lab\footnote{https://datalab.noao.edu/} with the same sky coverage and constraint parameters ($flags<4$ and $class\_star\geq0.9$). By crossmatching with our source catalog using a search radius of $1''$, we retrieved 132,342 measurements in $u$-band, 355,823 measurements in $g$-band, 497,443 measurements in $r$-band, 210,496 measurements in $i$-band, 404,087 measurements in $z$-band, and 16,916 measurements in Y-band. However, due to the irregular sampling in different filters, the observational timescales vary from tenths of a day to more than one thousand days with several to hundreds measurements. Thus, we required at least five valid measurements for each target in individual filter in order to calculate the MAD, SD and Amp. We need to emphasize that as a result of the irregular sampling, and particularly, a strong systematic effect we discovered during the data processing (instead of a relatively uniform distribution, different SMC regions show variable levels of average variabilities), therefore, the calculated variability may not fully represent the true variability of the target and the user should be cautious when using these values. The middle panel of Figure~\ref{optical_var_plot} shows the spatial distribution of targets with variability statistics in different filters, including 17,548 (38.60\%) targets in $u$-band, 25,897 (57.00\%) targets in $g$-band, 29,572 (65.04\%) targets in $r$-band, 19,632 (43.18\%) targets in $i$-band, 35,850 (78.85\%) targets in $z$-band, and 1,212 (2.67\%) targets in Y-band.

\begin{figure*}
\center
\includegraphics[bb=55 445 560 615]{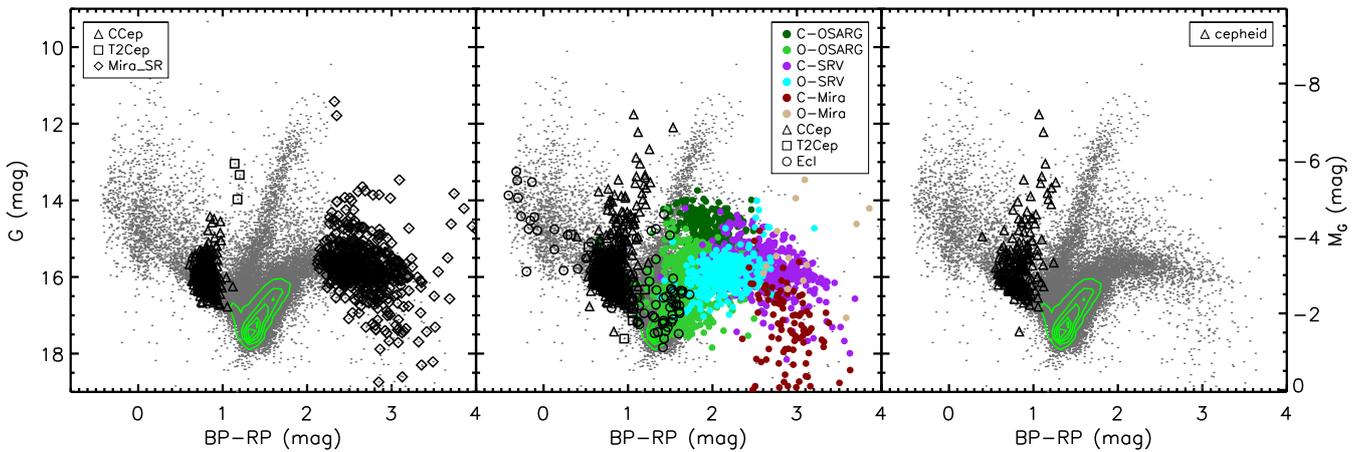}
\caption{Gaia color-magnitude diagrams with variable classifications from \textit{Gaia} (left), OGLE (middle), and VMC (right). Note that the left panel only shows targets with $best\_class\_score\geq0.5$ from \textit{Gaia}. Green contours represent the number density.  
\label{cmd_var_class}}
\end{figure*}

Finally, we obtained additional data from the Optical Gravitational Lensing Experiment (OGLE; \citealt{Udalski1992, Szymanski2005, Udalski2008, Udalski2015}) by using a search radius of $1''$ from both OGLE-\uppercase\expandafter{\romannumeral3} Catalog of Variable Stars (O$^3$CVS) and OGLE-\uppercase\expandafter{\romannumeral4} Collection of Variable Stars (O$^4$CVS), which resulted in 8,956 (19.70\%) Long Period Variables (LPVs) from O$^3$CVS, and 482 (1.06\%) Classical Cepheids (CCeps), 11 (0.02\%) Type \uppercase\expandafter{\romannumeral2} Cepheids (T2Ceps), and 87 (0.19\%) Eclipsing Binaries (Ecls) from O$^4$CVS \citep{Soszyski2011, Soszyski2015, Pawlak2016, Soszyski2018} as shown in the right panel of Figure~\ref{optical_var_plot}. However, since the OGLE data are calculated by using Fourier analysis, which is different from our dataset, only the classifications are used in the further analysis. Figure~\ref{cmd_var_class} shows \textit{Gaia} color-magnitude diagrams with variable classifications from \textit{Gaia} (left; only shows targets with $best\_class\_score\geq0.5$), OGLE (middle), and VMC (right).

We also checked the time-series data from SkyMapper DR1.1. However, this first data release provides only few measurements ($\sim$1-2) covering $\sim$300 days, which is not suitable for robust variability calculation.

Our SMC source catalog of 45,466 targets is available in its entirety in CDS. Table~\ref{isample} documents the columns found in the catalog. Targets without errors indicate either a 95\% confidence upper limit, or the errors are simply too large to be reliable (e.g., $>$1.0 mag). 


\section{Identifying Evolved Massive Star Candidates on the Color-Magnitude Diagrams}

As we focus on the evolved dusty massive stars, the primary task is to identify them. We utilize the evolutionary tracks and synthetic photometry from Modules for Experiments in Stellar Astrophysics (MESA; \citealt{Paxton2011, Paxton2013, Paxton2015, Paxton2018}) Isochrones \& Stellar Tracks (MIST\footnote{http://waps.cfa.harvard.edu/MIST/}; \citealt{Choi2016, Dotter2016}), which covers a wide range of ages, masses and metallicities by using solar-scaled abundance under a single computational framework, to identify evolved massive star candidates on the CMDs of our multiwavelength source catalog. 

We used a canonical value of 18.95 as the distance modulus of the SMC \citep{Graczyk2014, Scowcroft2016}. Since the metallicity of SMC is about 10\% to 20\% solar \citep{Russell1992, Dobbie2014, DOnghia2016}, we adopted the chemical composition of -1.0 to -0.7 dex for [Fe/H]. The non-rotation and rotation ($V/V_{crit}=0.40$) models of 7 to 40 $M_{\sun}$ were computed with no extinction and extinction of $A_V=1.0$ mag, respectively \citep{Cardelli1989, Zaritsky2002, Haschke2011, Gao2013}. We chose the color-magnitude combinations based on the available synthetic photometry in MIST with relative percentage $>90\%$ (see Table~\ref{fpercent}), at longer wavelengths (some dusty massive stars might not be identified in the shorter wavelengths due to higher extinction and reddening). In those models were also clearly shown the Yellow Void between blue supergiant stars (BSGs) and RSGs in order to identify YSGs. 

\begin{figure*}
\includegraphics[bb=55 405 560 650, scale=0.52]{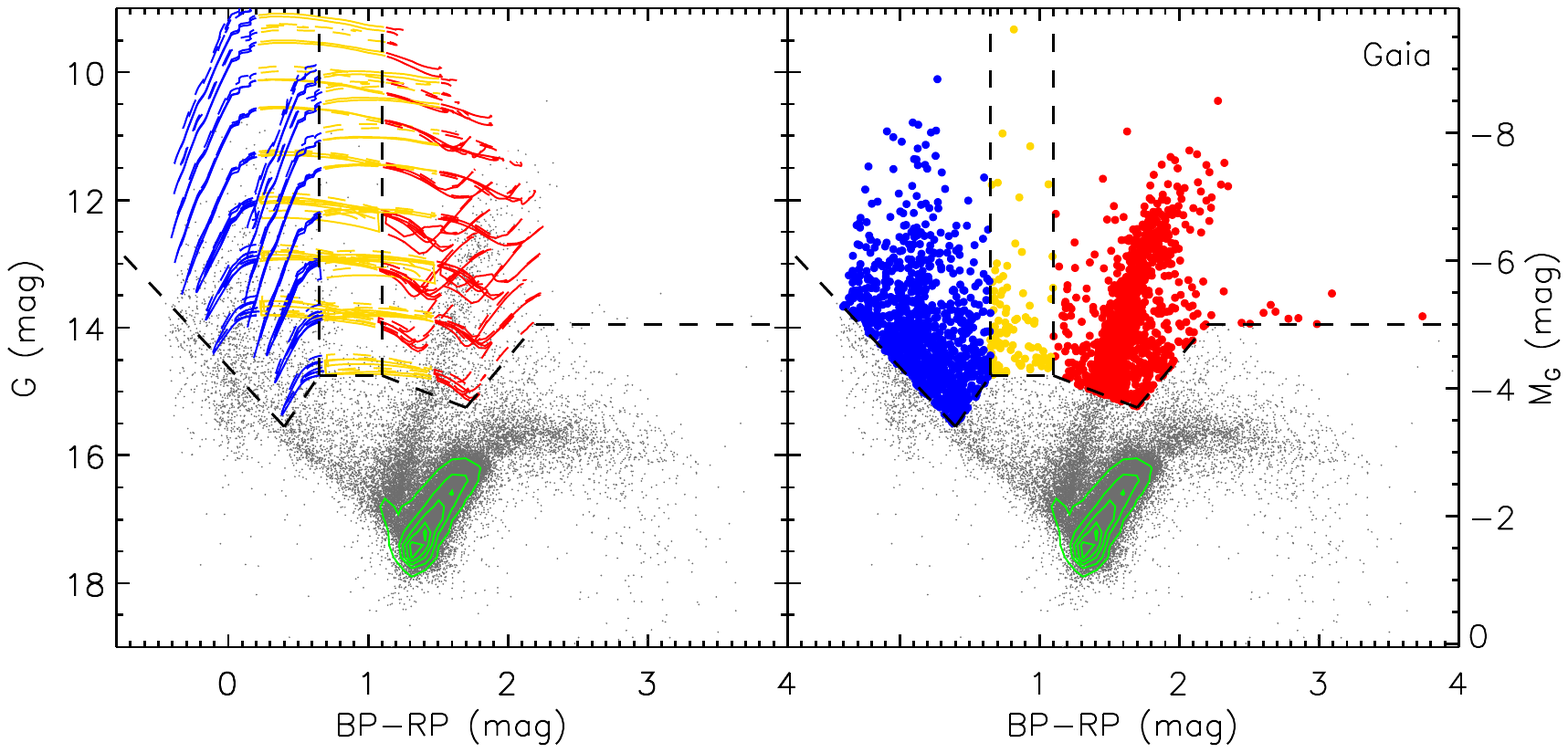}
\includegraphics[bb=55 405 560 650, scale=0.52]{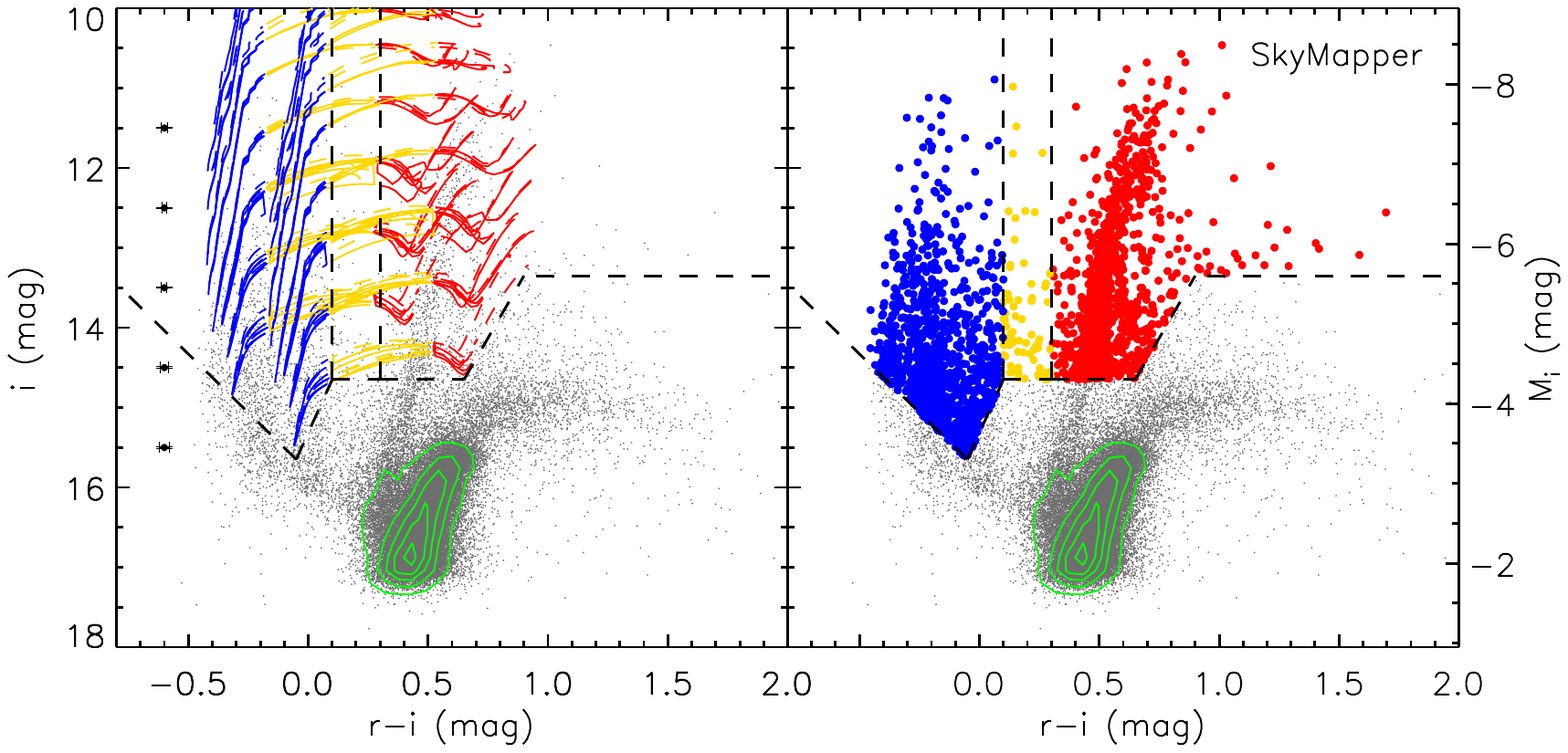}
\includegraphics[bb=55 405 560 650, scale=0.52]{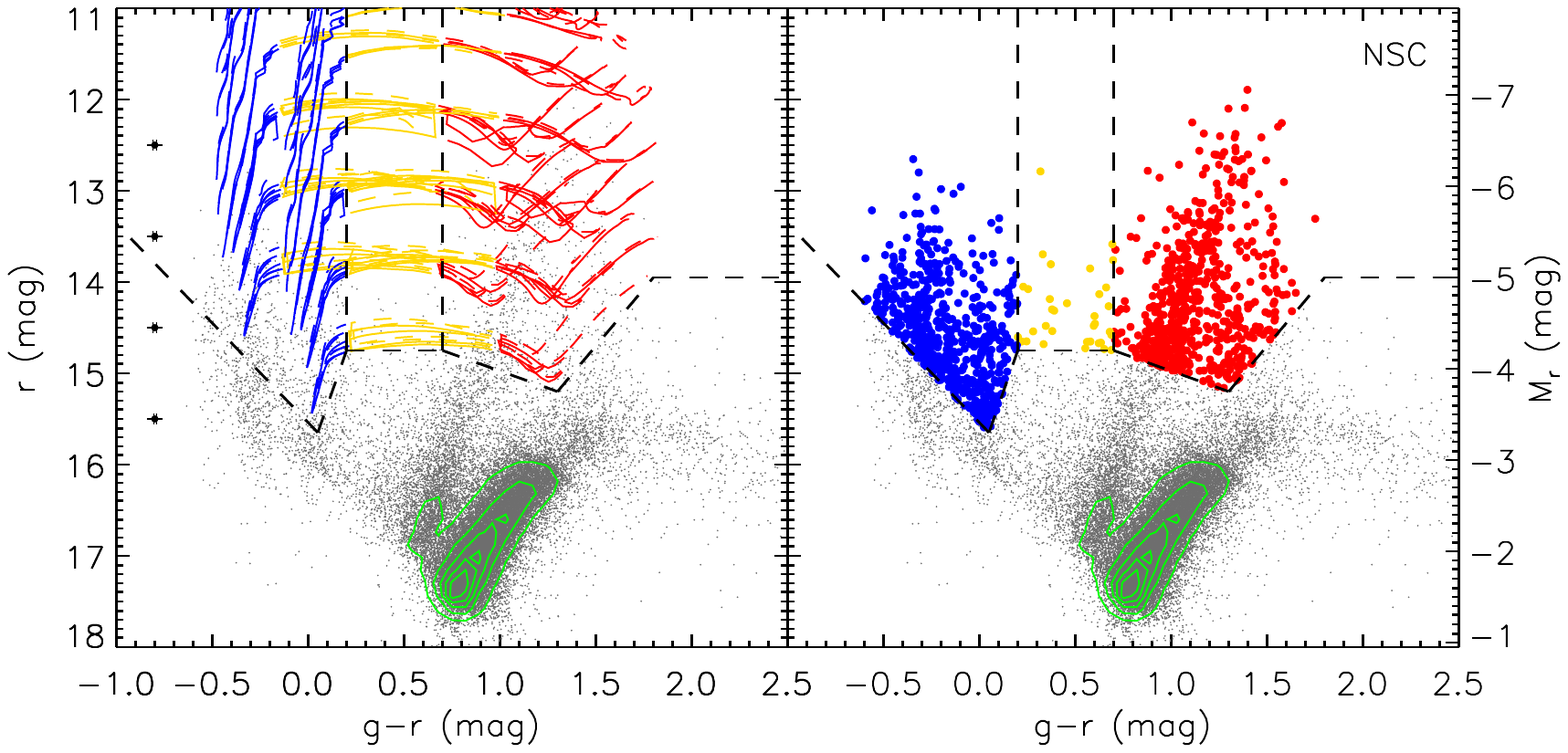}
\includegraphics[bb=55 405 560 650, scale=0.52]{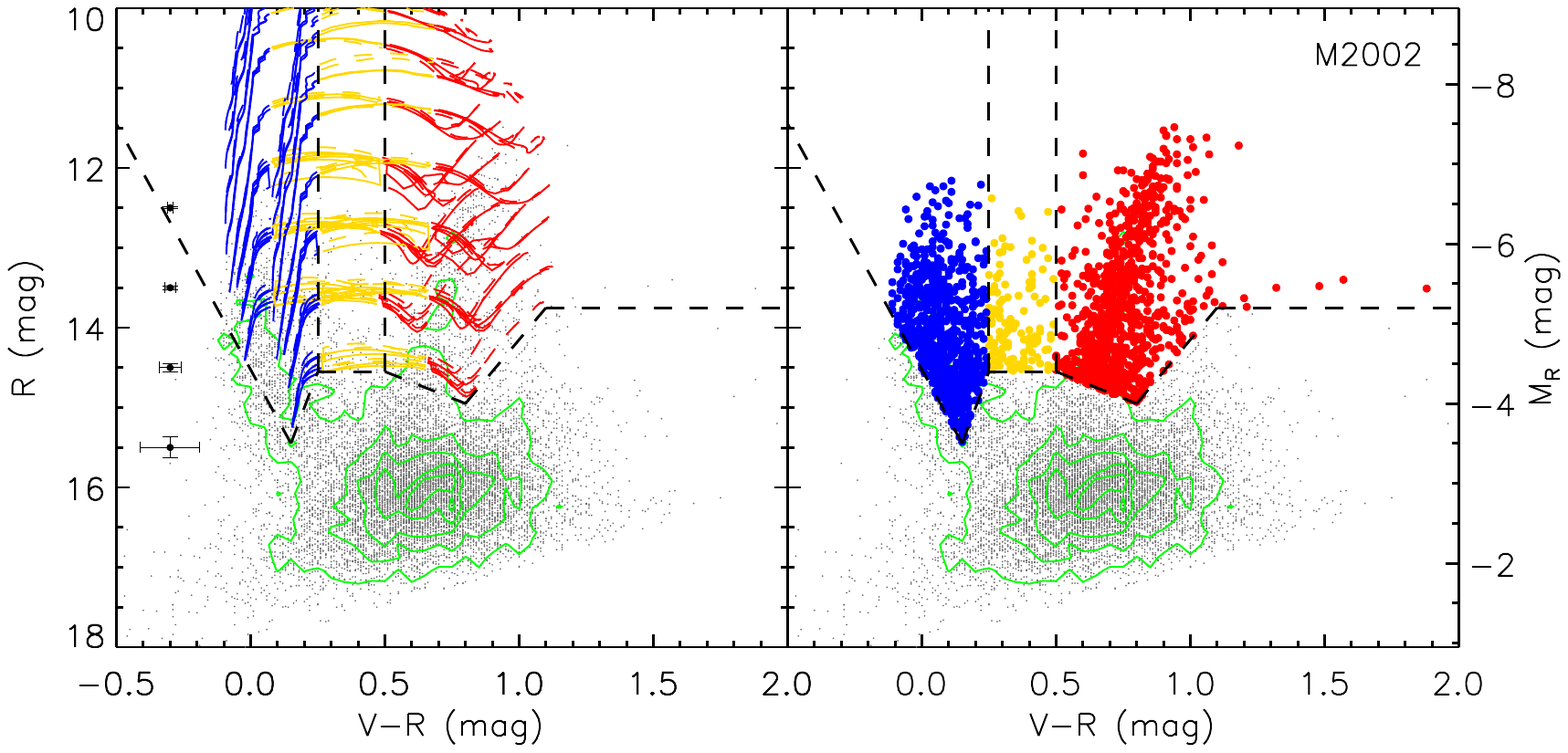}
\caption{Color-magnitude diagrams of \textit{Gaia} (upper left), SkyMapper (upper right), NSC (bottom left) and M2002 (bottom right) datasets. In each diagram, the left panel shows the CMD overlapped with MIST evolutionary tracks of 7, 9, 12, 15, 20, 25, 32 and 40 $M_{\sun}$ and color coded as BSG (blue), YSG (yellow) and RSG (red) phases. The regions of each type of evolved massive stars are outlined by the dashed lines with color and magnitude criteria listed in Table~\ref{outlines}. Targets without errors are not shown in the CMDs, while the average photometric uncertainties are indicated when available. The right panel shows the selected targets for each type of massive stars with the same color convention. The RSGs region is empirically extended from the reddest and faintest points of the models to even redder but not fainter area in order to avoid the contamination from x-AGBs. Green contours represent the number density. The diagrams show a clear bimodal distribution of the BSGs and RSGs candidates with few YSGs candidates lying between them. For the RSGs population, it reaches to approximately the TRGB without blending into AGBs, which is beyond the boundary of 7 $M_{\sun}$ shown by the model. See text for details. \label{mist_multi}}
\end{figure*}

Figure~\ref{mist_multi} shows multiple optical CMDs of \textit{Gaia}, SkyMapper, NSC and M2002 datasets, overlapped with the evolutionary tracks of 7, 9, 12, 15, 20, 25, 32, and 40 $M_{\sun}$ generated from MIST synthetic photometry. The tracks are color coded based on the equivalent evolutionary phases (EEPs) from core helium burning to carbon burning, and $T_{\rm eff}$ of $7500~K<T_{\rm eff}$ (blue; BSGs), $5000<T_{\rm eff}\leq7000~K$ (yellow; YSGs) and $T_{\rm eff}\leq5000~K$ (red; RSGs) \citep{Neugent2010}. The regions of each type of evolved massive stars are outlined by the dashed lines with color and magnitude criteria listed in Table~\ref{outlines}. Targets without errors are not shown in the CMDs, while the average photometric uncertainties are indicated when available. Due to the relatively large MLR during the RSGs phase, the star could be heavily obscured by the surrounding dust envelope \citep{Smith2001, Massey2005, Levesque2006, Yang2018, Ren2019}. Thus, we also empirically extended the RSGs region from the reddest and faintest points of the models to the even redder but not fainter area, in order to avoid the contamination from extreme AGBs (x-AGBs, stars are most likely to be experiencing a ``superwind'', where the mass-loss rate can increase by a factor of 10, and a thick dust envelope obscures the star at optical wavelengths; \citealt{vanLoon2006, Boyer2011}) as shown in the diagrams. It may be that some super-AGBs (stars in the mass range of $\sim7-10~M_{\sun}$ that represent a transition to the more massive supergiant stars and are characterised by degenerate off-centre carbon ignition analogous to the earlier helium flash; \citealt{Herwig2005, Siess2006, Siess2007, Siess2010, Groenewegen2009, Doherty2017}) are also selected by the extension. However, we expect that they can be rejected by using several methods as shown in \citet{Yang2018}. Moreover, inevitably, there will be contamination of the main sequence massive stars at the blue end, which cannot be easily disentangled. The diagrams show a clear bimodal distribution of the BSGs and RSGs candidates with few YSGs candidates lying between them. We notice that the conventional lower limit of the stellar mass for ``massive star'' is usually defined as eight solar mass. However, our initial tests show that the observational data are fitted better with the evolutionary track of seven solar mass (or even lower) as shown in the diagrams. This may be due to the treatment of parameters in the model such as convective overshooting, rotation, mixing, metallicity, and MLR, or the uncertainties of extinction and/or bolometric correction, or it may be true that ``massive star'' do have lower limit of stellar mass, which are complicated and need further investigation.

\begin{figure*}
\center
\includegraphics[bb=55 405 560 650, scale=0.52]{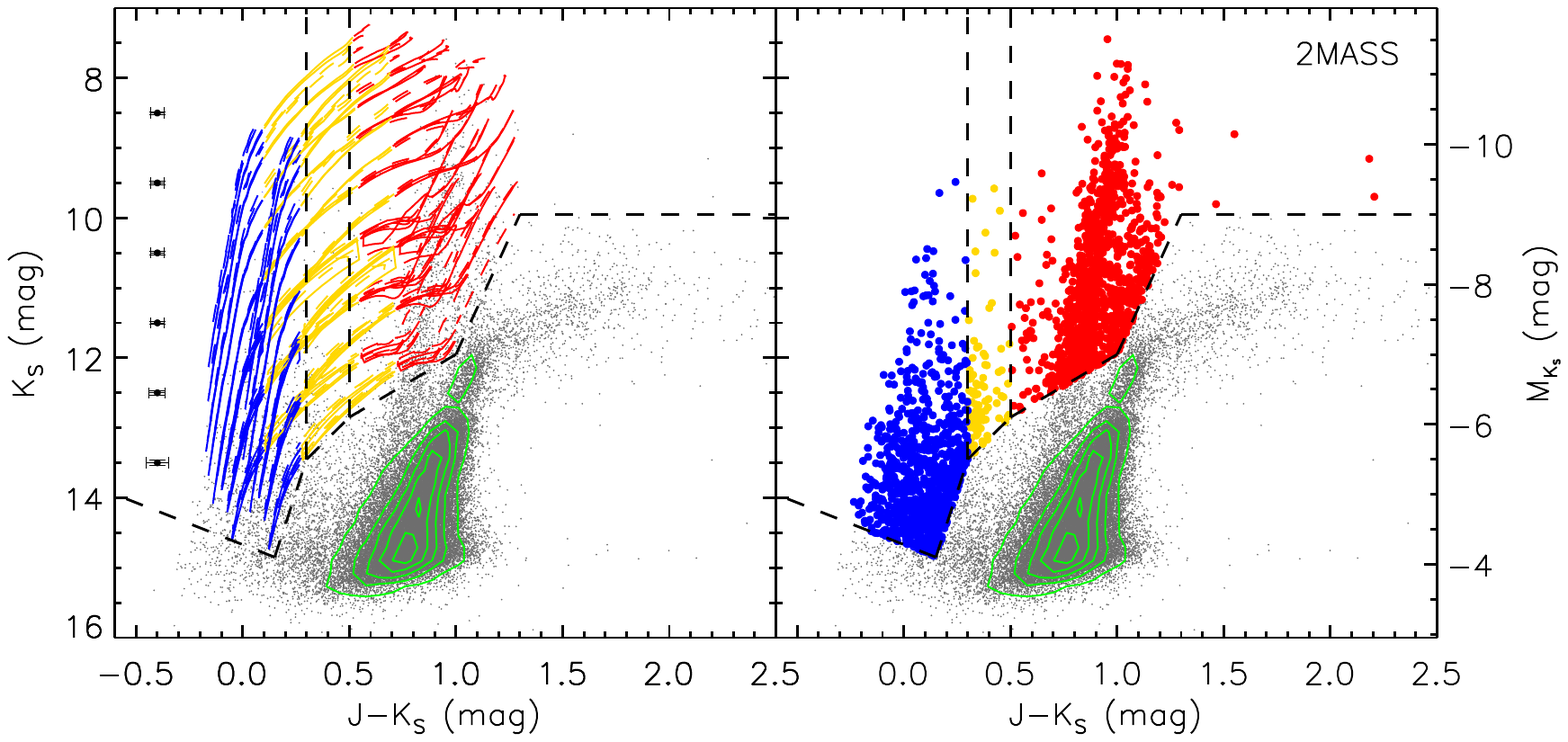}
\includegraphics[bb=160 405 455 650, scale=0.52]{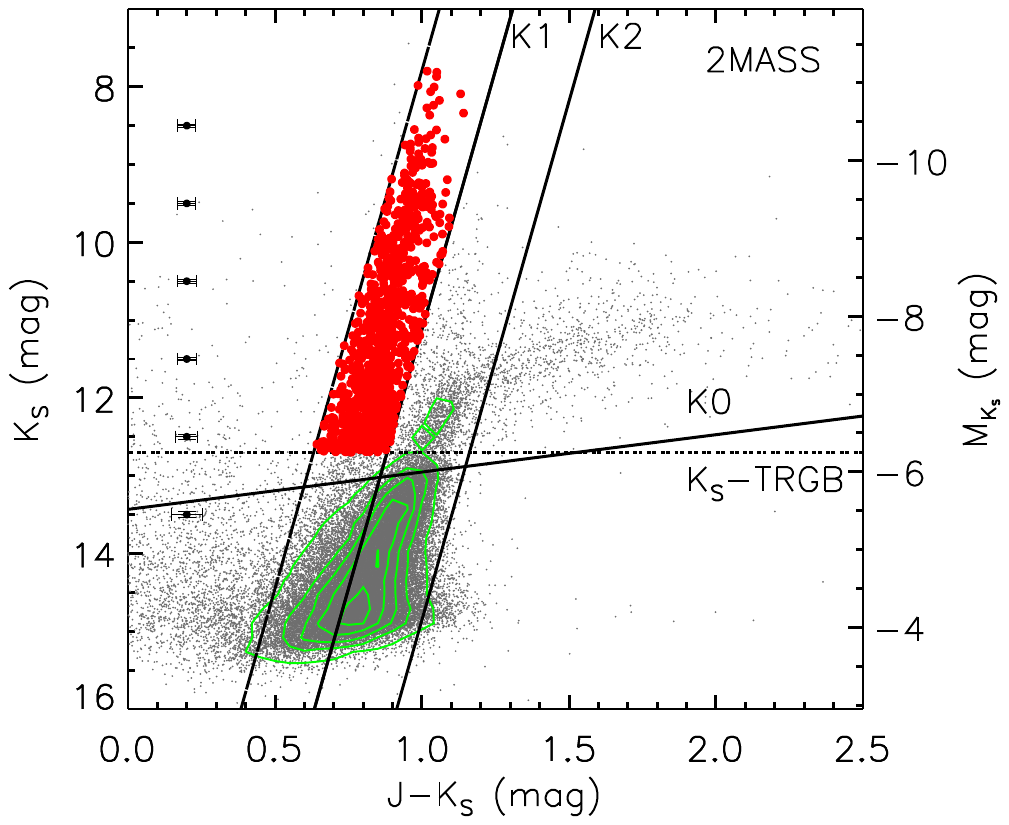}
\caption{K$_{\rm S}$ versus $\rm J-K_S$ diagram for the 2MASS dataset. The left panel is similar to Figure~\ref{mist_multi} as massive star candidates selected by MIST tracks. The right panel shows the definitions of C-AGBs, O-AGBs, x-AGBs and RSGs regions by using the theoretical $\rm J-K_S$ color cuts from \citet{Cioni2006a} and \citet{Boyer2011}. Targets selected as RSGs candidates are shown as red color. Green contours represent the number density.  \label{nir_multi}}
\end{figure*}

It can be seen from the Figure~\ref{mist_multi} that in the optical bands, generally, massive stars evolve horizontally across the upper part of CMD, while it is slightly different in the NIR bands. The left panel of Figure~\ref{nir_multi} shows the 2MASS CMD overlapped with MIST tracks, for which the tracks extend from the fainter and bluer region to the brighter and redder area due to the displacement of $T_{\rm eff}$ and intrinsic emission peaks. The right panel of Figure~\ref{nir_multi} shows a different method to classify RSGs originated from \citet{Cioni2006a} and \citet{Boyer2011} (hereafter CB method), where Oxygen-rich AGBs (O-AGBs) are defined by K$_{\rm S}$ $<$ K$_{\rm S}$-band TRGB (K$_{\rm S}$-TRGB$\approx$12.7 mag) and K1 $<\rm J-K_{S} <$ K2, Carbon-rich AGBs (C-AGBs) are defined by K$_{\rm S}$ $<$ K$_{\rm S}$-TRGB and K2 $<\rm J-K_{S} <$ 2.1 mag, x-AGBs are defined by $\rm J-K_{S}>2.1$ mag (all AGB stars are brighter than the K0 line, except x-AGBs), and RSGs are defined by $\rm \Delta(J-K_{S})=0.25$ mag from the O-AGBs shown as the dashed line in the diagram (the distance and 0.05 mag for the metallicity between LMC and SMC are corrected; \citealt{Cioni2006b}). It occurs to us that for the CB method, it covers almost the whole magnitude range of RSGs population down to the K$_{\rm S}$-TRGB, where the MIST tracks are unable to cover. However, MIST tracks are more broadened with a part of the tracks also extending to the O-AGBs region. This discrepancy between MIST tracks and CB method will be addressed in our next paper. A simple calculation by using a constant bolometric correction (BC) of $\rm BC_{K_S}=2.69$ \citep{Davies2013} and $\rm A_{K_S}=0.1$ mag shows that the RSGs candidates close to K$_{\rm S}$-TRGB (e.g., $\rm K_S=12.6$ mag) only correspond to $\sim10^{3.4}$ ($\sim$2,500) solar luminosity ($L_\sun$). 

\begin{figure*}
\center
\includegraphics[bb=90 365 515 715]{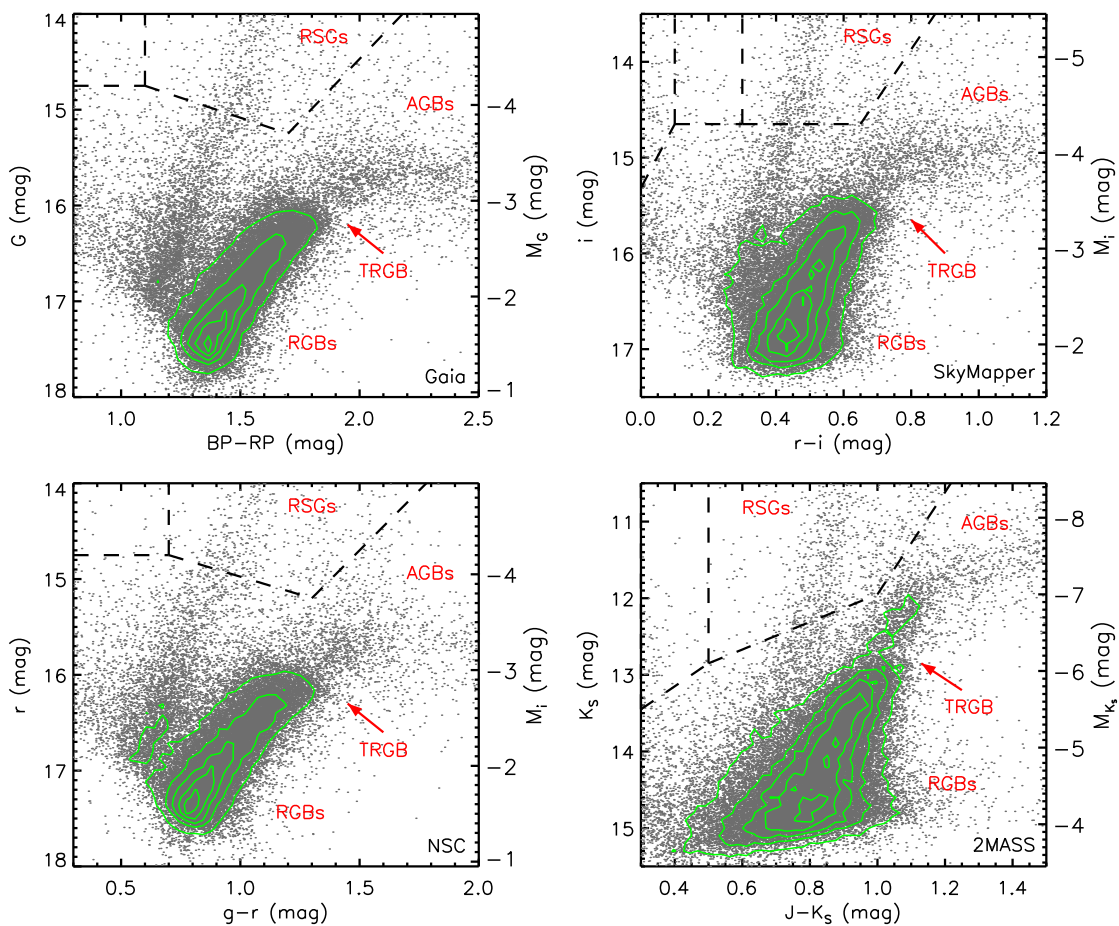}
\caption{Zoom in regions of $\rm 0.8<BP-RP<2.5$ and $\rm 14.0<G<18.0$ for \textit{Gaia} (upper left), $\rm 0<r-i<1.2$ and $\rm 13.5<i<17.5$ for SkyMapper (upper right), $\rm 0.3<g-r<2.0$ and $\rm 14.0<r<18.0$ for NSC (bottom left), and $\rm 0.3<J-K_S<1.5$ and $\rm 10.5<K_S<15.5$ for 2MASS (bottom right). The populations of RSGs, AGBs, RGBs, and the TRGB are indicated on the diagram. It can be seen that the separation between the RSG and AGB populations is quite clear, even at faint magnitude. \label{agb_rsg}}
\end{figure*}

We also notice one important observational piece of evidence that, for the RSGs population, there is a distinct branch stretching continuously from the top of the luminous cool region towards the relatively faint warm area, reaching approximately to the tip of the red giant branch (TRGB) without blending into the AGBs population. It is obviously beyond the limit of 7 $M_{\sun}$ track and most probably reaching down to 6 $M_{\sun}$. False detection is ruled out due to two reasons. Firstly, all the \textit{Gaia}, SkyMapper, NSC and 2MASS data show the same tendency except the M2002 data, which is likely due to the photometric sensitivity. Secondly, our sample is strictly constrained by the \textit{Gaia} astrometric solution and verified by comparing with the result of \citet{Gaia2018b}, which indicate that the luminosities and colors of our targets are almost certain and the photometric errors are also very small at the magnitude range around the TRGB. Figure~\ref{agb_rsg} shows the zoom in regions of $\rm 0.8<BP-RP<2.5$ and $\rm 14.0<G<18.0$ for \textit{Gaia} (upper left), $\rm 0<r-i<1.2$ and $\rm 13.5<i<17.5$ for SkyMapper (upper right), $\rm 0.3<g-r<2.0$ and $\rm 14.0<r<18.0$ for NSC (bottom left), and $\rm 0.3<J-K_S<1.5$ and $\rm 10.5<K_S<15.5$ for 2MASS (bottom right). From the diagram and the observational point of view, there is no doubt that the separation between the RSG and AGB populations is quite clear, even at faint magnitude. Additionally, the distribution of a large sample of spectroscopic SMC RSGs (more than 300 and brighter than $\rm K_S\approx11.0~mag$) on the 2MASS CMD follows almost exactly the MIST tracks, indicating the excellence of the MIST model prediction (details will be presented in our next paper). Thus, even though there may still be inevitable contamination of AGBs at the faint magnitude end of the RSG sample, we do believe that it should be small, and will be easily resolved by the next generation of large-scale spectroscopic surveys (e.g., 4MOST, MOONS; \citealt{Cirasuolo2012, deJong2012}). Despite of the small contamination, the overlapping of lower (initial) masses of RSGs and the upper limits of AGBs, and the clear separation between them at faint magnitude range are also very interesting and puzzling. It means that, at given magnitude, some stars evolve to RSGs and the others evolve to AGBs. Currently, we have no certain explanation for this. It may be due to the rotation, binarity, chemical composition, MLR, or even more sophisticated evolution. Nevertheless, this may indicate the uniqueness of RSGs population, which connects the evolved massive and intermediate stars, since stars with initial mass around 6 to 8 $M_{\sun}$ are thought to go through a second dredge-up to become AGBs \citep{Eldridge2007}. The low mass RSGs may be also related to the intermediate luminosity optical transients (ILOTs; \citealt{Prieto2008, Bond2009, Berger2009}). Still, more investigations are needed to confirm the true nature of RSGs.

\begin{figure*}
\center
\includegraphics[bb=60 425 550 635, scale=1.05]{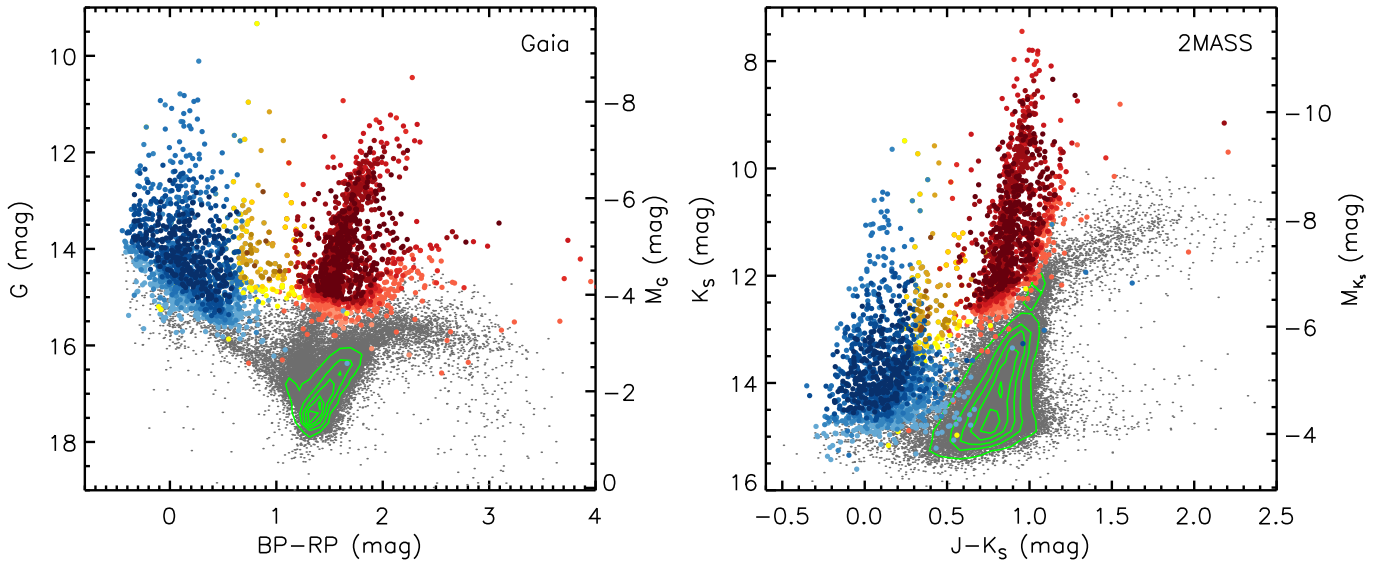}
\caption{Color-magnitude diagrams of \textit{Gaia} (left) and 2MASS (right) with RSG (red), YSG (yellow), and BSG (blue) candidates overlapped, where the colors are coded from dark (Rank 0) to light (Rank 5) based on the ranks. The RSG branch extends towards fainter magnitudes with few candidates scattered in the much fainter and redder region in the optical band, which is likely caused by the circumstellar dust envelope. Green contours represent the number density. 
\label{cmd_candidates_rank}}
\end{figure*}

\begin{figure}
\center
\includegraphics[bb=115 365 460 690, scale=0.7]{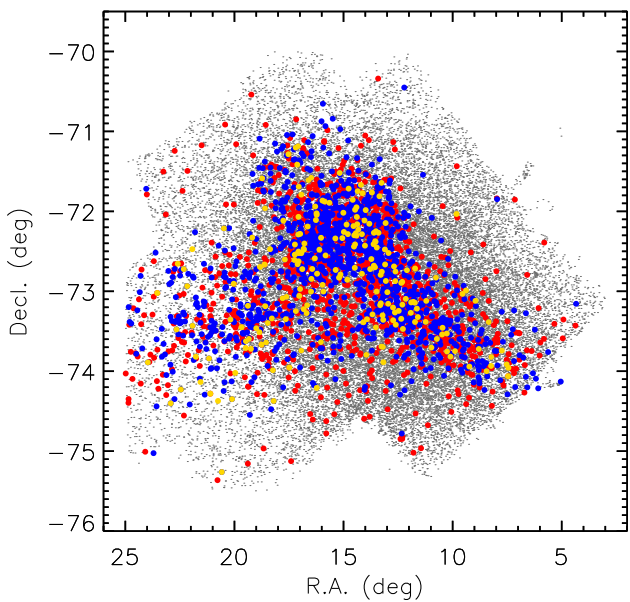}
\caption{Spatial distribution of evolved massive star candidates. Due to the interaction between LMC and SMC, a bunch of candidates are stretched towards the MB. 
 \label{candi_spatial}}
\end{figure}

We combined the candidates of each type of evolved massive stars from different datasets and removed the duplications, which resulted in 1,405 RSG, 217 YSG and 1,369 BSG candidates listed in Table~\ref{model_ident}. Since the candidates were mostly selected based on the MIST model prediction in different datasets with variety of filters, sky coverage, photometric sensitivities and qualities, it was difficult to judge how reliable the candidacies were. Therefore, we ranked (Rank 0 to 5) the candidates based on the intersection between different CMDs, where Rank 0 indicated that a target was identified as the same type of evolved massive star in all five datasets (\textit{Gaia}, SkyMapper, NSC, M2002 and 2MASS) by the MIST models and so on, and Rank 5 indicated the additional RSG candidates identified by the CB method but not recovered by the MIST models. Table~\ref{rank_number} shows the numbers of ranked candidates for each type of evolved massive stars. Figure~\ref{cmd_candidates_rank} illustrates two CMDs (\textit{Gaia} and 2MASS) for all the candidates with ranks, where RSG, YSG, and BSG candidates are color coded in red, yellow, and blue colors ranging from dark (Rank 0) to light (Rank 5). Detailed information about each type of evolved massive star candidates are presented in separate tables available in CDS with the similar formats as Table~\ref{isample}, except the Rank has been added as an additional column in the end of each table. It has to be emphasized that some candidates may have different classifications at the same time, which is likely due to either the inevitable slightly overlapping of adjacent types of massive stars, or the larger uncertainties of photometry at the fainter magnitudes. Moreover, there are very few stray candidates way into other populations deserved to be investigated. For example, a couple of BSG candidates appearing in the RSGs region in the 2MASS CMD are likely due to their thick circumstellar dust envelope, since they are also very bright in the MIR bands. Similarly, we notice that there are few RSG candidates scattered in the much fainter and redder regions in the \textit{Gaia} CMD. However, simultaneous inspection of the optical and MIR CMDs shows that almost all the scattered candidates in the bottom right region show infrared excess and/or high MIR luminosity, which may indicate that the dimming in the optical band is caused by the circumstellar dust envelope. Finally, Figure~\ref{candi_spatial} shows the spatial distribution of evolved massive star candidates. It is clearly shown that due to the interaction between LMC and SMC, a bunch of candidates are stretched towards the MB. These populations follow the distribution of star formation in the SMC which is predominant along the bar of the SMC and extend to the LMC through the MB. This is depicted from the star cluster (especially for those with ages $<$100~Myr) distribution of the SMC \citep{Bitsakis2018}. Further detailed analysis of identified massive star populations will be presented in our following papers.

Finally, we would like to emphasize again one important thing that, our purpose was to study the evolved dusty massive stars in the SMC by identifying BSG, YSG, and RSG populations primarily based on the infrared detections, and constrained by the astrometric solutions and the evolutionary models. In that sense, our sample is \textit{NOT} complete due to the strict constraints on the astromety, deblending, limitations of the models, and more important, the (presumably) serious differences in the luminosity completeness limits as a function of temperature. As we adopted the magnitude cut at the MIR wavelength (IRAC1 or WISE1 $\leq$ 15.0 mag), hotter stars would suffer more from incompleteness due to the weaker radiation at the far end of the Rayleigh-Jeans tail than the cooler stars. Even for the hot stars themselves, the degeneracy in the infrared detection would occur for hotter stars with larger MLR (less contribution from the stellar emission and more contribution from the dust emission) and cooler stars with smaller MLR (vice versa), since they might have similar radiative intensities at the same MIR wavelengths, which also happens in all magnitude ranges (not to mention the effect of binarity). One should \textit{NOT} simply conclude that the BSGs to RSGs ratio (B/R ratio) is about 1:1 in the SMC, since both the observation and theoretical prediction indicates that there will be more BSGs than the RSGs (e.g., B/R ratio$\sim$4 or more) in the SMC \citep{Meylan1982, Humphreys1984, Guo2002}.

\section{Summary}

We present a clean, magnitude-limited (IRAC1 or WISE1 $\leq$ 15.0 mag) multiwavelength source catalog for the SMC with 45,466 targets in total. We intend to build our catalog as a comprehensive dataset serving as an anchor for future studies, especially for massive star populations at low-metallicity. The catalog contains data in 50 different bands including 21 optical and 29 infrared bands, retrieved from SEIP, VMC, IRSF, \textit{AKARI}, HERITAGE, \textit{Gaia}, SkyMapper, NSC, M2002, and \textit{GALEX} datasets, ranging from ultraviolet to far-infrared. Additionally, radial velocities and spectral classifications were collected from the literature, as well as the infrared variability statistics, including MAD, SD, and Amp, derived from \textit{WISE}, SAGE-Var, VMC and IRSF, and the optical variability statistics derived from \textit{Gaia}, NSC and OGLE. 

The catalog was essentially built upon a $1''$ crossmatching and a $3''$ deblending between SEIP source list and \textit{Gaia} photometric data. We further constrained the PMs and parallaxes from \textit{Gaia} DR2 to remove the foreground contamination, by applying a Gaussian profile in parallax with an additional elliptical constraint derived from $PM_{R.A.}$ and $PM_{Decl.}$. We estimated that about 99.5\% of the targets in our catalog were most likely to be genuine members of the SMC. 

By using the evolutionary tracks and synthetic photometry from MIST and also the theoretical $\rm J-K_S$ color cuts from CB method, we identified three evolved massive star populations in the SMC, namely the BSGs, YSGs and RSGs, in five different CMDs. There are 1,405 RSG, 217 YSG and 1,369 BSG candidates, respectively. We emphasized that our sample was \textit{NOT} complete due to several reasons. We ranked the candidates based on the intersection of different CMDs, where the source with the most intersections was given the highest rank. A comparison between the models and observational data shows that, the lower limit of initial mass for the RSGs population may reach down to 7 or even 6 $M_{\sun}$ and the RSG is well separated from the AGB population even at faint magnitude, making RSGs a unique population connecting the evolved massive and intermediate stars, since stars with initial mass around 6 to 8 $M_{\sun}$ are thought to go through a second dredge-up to become AGBs. 

We encourage the interested reader to further exploit the potential of our catalog, including, but not limited to, massive stars, supernova progenitors, star formation history, stellar population, stellar kinematics, chemical evolution, individual/integrated spectral energy distribution, time-domain astronomy, and so on. A further detailed analysis of the RSG population in the SMC will be presented in our next paper.

\section{Acknowledgments}

We would like to thank the anonymous referee for many constructive comments and suggestions. We acknowledge funding from the European Research Council (ERC) under the European Union's Horizon 2020 research and innovation programme (grant agreement number 772086), and from Hubble Catalog of Variables project funded by the European Space Agency (ESA) under contract No.4000112940. B.W.J. and J.G. gratefully acknowledge support from the National Natural Science Foundation of China (Grant No.11533002 and U1631104). We thank Man I Lam and Stephen A. S. de Wit for helpful comments and suggestions.

This publication makes use of data products from the Two Micron All Sky Survey, which is a joint project of the University of Massachusetts and the Infrared Processing and Analysis Center/California Institute of Technology, funded by the National Aeronautics and Space Administration and the National Science Foundation. This work is based in part on observations made with the \textit{Spitzer} Space Telescope, which is operated by the Jet Propulsion Laboratory, California Institute of Technology under a contract with NASA. This publication makes use of data products from the Wide-field Infrared Survey Explorer, which is a joint project of the University of California, Los Angeles, and the Jet Propulsion Laboratory/California Institute of Technology. It is funded by the National Aeronautics and Space Administration. This publication makes use of data products from the Near-Earth Object Wide-field Infrared Survey Explorer (NEOWISE), which is a project of the Jet Propulsion Laboratory/California Institute of Technology. NEOWISE is funded by the National Aeronautics and Space Administration. This research has made use of the NASA/IPAC Infrared Science Archive, which is operated by the Jet Propulsion Laboratory, California Institute of Technology, under contract with the National Aeronautics and Space Administration.

This work has made use of data from the European Space Agency (ESA) mission {\it Gaia} (\url{https://www.cosmos.esa.int/gaia}), processed by the {\it Gaia} Data Processing and Analysis Consortium (DPAC, \url{https://www.cosmos.esa.int/web/gaia/dpac/consortium}). Funding for the DPAC has been provided by national institutions, in particular the institutions participating in the {\it Gaia} Multilateral Agreement.

This research uses services or data provided by the NOAO Data Lab. NOAO is operated by the Association of Universities for Research in Astronomy (AURA), Inc. under a cooperative agreement with the National Science Foundation.

The national facility capability for SkyMapper has been funded through ARC LIEF grant LE130100104 from the Australian Research Council, awarded to the University of Sydney, the Australian National University, Swinburne University of Technology, the University of Queensland, the University of Western Australia, the University of Melbourne, Curtin University of Technology, Monash University and the Australian Astronomical Observatory. SkyMapper is owned and operated by The Australian National University's Research School of Astronomy and Astrophysics. The survey data were processed and provided by the SkyMapper Team at ANU. The SkyMapper node of the All-Sky Virtual Observatory (ASVO) is hosted at the National Computational Infrastructure (NCI). Development and support the SkyMapper node of the ASVO has been funded in part by Astronomy Australia Limited (AAL) and the Australian Government through the Commonwealth's Education Investment Fund (EIF) and National Collaborative Research Infrastructure Strategy (NCRIS), particularly the National eResearch Collaboration Tools and Resources (NeCTAR) and the Australian National Data Service Projects (ANDS).

This research has made use of the SIMBAD database and VizieR catalog access tool, operated at CDS, Strasbourg, France, and the Tool for OPerations on Catalogues And Tables (TOPCAT; \citealt{Taylor2005}).

Based on data products from observations made with ESO Telescopes at the La Silla or Paranal Observatories under ESO programme ID 179.B-2003.

\begin{table*}
\caption{Number of Detected Targets in Each Filter of the SMC Source Catalog}
\label{fpercent}
\begin{tabular}{cccc}
\toprule\toprule
Filter & $Number_{detected}$ & Absolute percentage & Relative percentage\tablefootmark{a} \\
\midrule
2MASS\_J & 45335 & 99.71\% & 100\% \\
2MASS\_H & 45335 & 99.71\% & 100\% \\
2MASS\_$\rm K_S$ & 45335 & 99.71\% & 100\% \\
IRAC1 & 44712 & 98.34\% & 99.47\% \\
IRAC2 & 44948 & 98.86\% & 100\% \\
IRAC3 & 38272 & 84.18\% & 85.15\% \\
IRAC4 & 35133 & 77.23\% & 18.16\% \\
MIPS24 & 1261 & 2.77\% & 2.81\% \\
WISE1 & 42460 & 93.39\% & 100\% \\
WISE2 & 42460 & 93.39\% & 100\% \\
WISE3 & 42460 & 93.39\% & 100\% \\
WISE4 & 42460 & 93.39\% & 100\% \\
Gaia\_G & 45466 & 100\% & 100\% \\
Gaia\_BP & 45466 & 100\% & 100\% \\
Gaia\_RP & 45466 & 100\% & 100\% \\
VMC\_Y & 18637 & 40.99\% & 99.99\% \\
VMC\_J & 18639 & 41.00\% & 100\% \\
VMC\_$\rm K_S$ & 18638 & 40.99\% & 99.99\% \\
IRSF\_J & 28645 & 63.00\% & 100\% \\
IRSF\_H & 28592 & 62.89\% & 99.81\% \\
IRSF\_$\rm K_S$ & 28513 & 62.71\% & 99.54\% \\
AKARI\_N3 & 555 & 1.22\% & 100\% \\
AKARI\_N4 & 551 & 1.21\% & 99.28\% \\
AKARI\_S7 & 274 & 0.60\% & 49.37\% \\
AKARI\_S11 & 134 & 0.29\% & 24.14\% \\
AKARI\_L15 & 40 & 0.09\% & 7.21\% \\
AKARI\_L24 & 18 & 0.04\% & 3.24\% \\
HERITAGE\_f70 & 1 & --- & --- \\
HERITAGE\_f100 & 1 & --- & --- \\
HERITAGE\_f160 & 2 & --- & --- \\
HERITAGE\_f250 & 2 & --- & --- \\
HERITAGE\_f350 & 2 & --- & --- \\
SkyMapper\_$u$ & 3185 & 7.01\% & 7.89\% \\
SkyMapper\_$v$ & 3711 & 8.16\% & 9.20\% \\
SkyMapper\_$g$ & 37949 & 83.47\% & 94.04\% \\
SkyMapper\_$r$ & 39930 & 87.82\% & 98.94\% \\
SkyMapper\_$i$ & 40356 & 88.76\% & 100\% \\
SkyMapper\_$z$ & 40171 & 88.35\% & 99.54\% \\
NSC\_$u$ & 26425 & 58.12\% & 68.24\% \\
NSC\_$g$ & 36700 & 80.72\% & 94.77\% \\
NSC\_$r$ & 37025 & 81.43\% & 95.61\% \\
NSC\_$i$ & 36297 & 79.83\% & 93.73\% \\
NSC\_$z$ & 38724 & 85.17\% & 100\% \\
NSC\_Y & 1250 & 2.75\% & 3.22\% \\
M2002\_V & 11630 & 25.58\% & 100\% \\
M2002\_B-V (B) & 11630 & 25.58\% & 100\% \\
M2002\_U-B (U) & 4000 & 8.80\% & 34.39\% \\
M2002\_V-R (R) & 11563 & 25.43\% & 99.42\% \\
GALEX\_FUV & 95 & 0.21\% & 57.93\% \\
GALEX\_NUV & 164 & 0.36\% & 100\% \\

\midrule
\end{tabular}
\tablefoot{
\tablefoottext{a}{Relative to the filter with the most matches in each dataset.}
}
\end{table*}

\begin{table*}
\caption{Observation Epochs of ALLWISE and NEOWISE-R} 
\label{epochtbl}
\begin{tabular}{cc}
\toprule\toprule
Beginning & Ending \\
(MJD-54000) & (MJD-54000) \\
\midrule
ALLWISE & \\
\midrule
 & 1400 \\
1400 & 1600 \\
\midrule
NEOWISE-R & \\
\midrule
2700 & 2870 \\
2870 & 3050 \\
3050 & 3230 \\
3230 & 3405 \\
3405 & 3600 \\
3600 & 3775 \\
3775 & 3955 \\
3955 & 4135 \\
4135 & 4325 \\
4325 &  \\
\midrule
\end{tabular}
\end{table*}

\begin{table*}
\footnotesize
\caption{SMC Source Catalog Contents}
\label{isample}
\centering
\begin{tabular}{lll}
\toprule\toprule
Column  & Name                                		& Description \\
1       & ID                                  		& Index \\
2       & R.A.(J2000)                         		& Right Ascension, J2000 (deg) \\
3       & Decl.(J2000)                        		& Declination, J2000 (deg) \\
4-9     & 2MASS\_J, ..., e\_2MASS\_$\rm K_S$      		& 2MASS photometries with uncertainties (mag) \\
10-19   & IRAC1, ..., e\_MIPS24               		& \textit{Spitzer} photometries with uncertainties (mag) \\
20-27   & WISE1, ..., e\_WISE4                		& \textit{WISE} photometries with uncertainties (mag) \\
28-33   & Gaia\_parallax, ..., e\_Gaia\_pmdec 		& \textit{Gaia} DR2 astrometric solutions with uncertainties (mas; mas/yr) \\
34-36   & Gaia\_G, ..., Gaia\_RP         		    & \textit{Gaia} DR2 mean magnitudes (mag) \\
37-38   & Gaia\_RV, e\_Gaia\_RV               		& \textit{Gaia} DR2 radial velocity with uncertainty (km/s) \\
39-44   & VMC\_Y, ..., e\_VMC\_$\rm K_S$          		& VMC photometries with uncertainties (mag) \\
45-50   & IRSF\_J, ..., e\_IRSF\_$\rm K_S$              & IRSF photometries with uncertainties (mag) \\
51-62   & AKARI\_N3, ..., e\_AKARI\_L24             & AKARI photometries with uncertainties (mag) \\
63-72   & HERITAGE\_f70, ..., e\_HERITAGE\_f350     & HERITAGE fluxes with uncertainties (mJy) \\
73-84   & SkyMapper\_$u$, ..., e\_SkyMapper\_$z$        & SkyMapper photometries with uncertainties (mag) \\
85-96   & NSC\_$u$, ..., e\_NSC\_Y 		            & NSC photometries with uncertainties (mag) \\
97-104  & M2002\_V, ..., e\_M2002\_V\_R 		    & M2002 photometries with uncertainties (mag) \\
105-108 & GALEX\_FUV, ..., e\_GALEX\_NUV 		    & GALEX photometries with uncertainties (mag) \\
109     & B2011\_SAGE\_Class                        & IR color classifications from \citet{Boyer2011} \\
110     & S2013\_YSO                                & YSO classifications from \citet{Sewilo2013} \\
111     & R2015\_IRS\_Class                         & MIR spectral classifications from \citet{Ruffle2015} \\
112     & B2010\_MK\_Class                          & Optical spectral classifications from \citet{Bonanos2010} \\
113-115 & GF2015\_MK\_SpVar, ..., GF2015\_MK\_Class & Optical spectral classifications from \citet{Gonzalez2015} (km/s) \\
116-117 & N2018\_RV, N2018\_MK\_Class               & Optical spectral classifications from \citet{Neugent2018} (km/s) \\
118-121 & Simbad\_RV, ..., Simbad\_Other\_Types     & Simbad classifications (km/s) \\
122-127 & Median\_WISE1, ..., Amp\_WISE1            & WISE1-band variability statistics (mag) \\
128-133 & Median\_WISE2, ..., Amp\_WISE2            & WISE2-band variability statistics (mag) \\
134-137 & Median\_IRAC1, ..., Amp\_IRAC1            & IRAC1-band variability statistics (mag) \\
138-141 & Median\_IRAC2, ..., Amp\_IRAC2            & IRAC2-band variability statistics (mag) \\
142-145 & Median\_VMC\_Y, ..., Amp\_VMC\_Y          & VMC Y-band variability statistics (mag) \\
146-149 & Median\_VMC\_J, ..., Amp\_VMC\_J          & VMC J-band variability statistics (mag) \\
150-153 & Median\_VMC\_$\rm K_S$, ..., Amp\_VMC\_$\rm K_S$  & VMC $\rm K_S$-band variability statistics (mag) \\
154     & VMC\_Type                                 & VMC variable classifications \\
155-158 & Median\_IRSF\_J, ..., Amp\_IRSF\_J        & IRSF J-band variability statistics (mag) \\
159-162 & Median\_IRSF\_H, ..., Amp\_IRSF\_H        & IRSF H-band variability statistics (mag) \\
163-166 & Median\_IRSF\_$\rm K_S$, ..., Amp\_IRSF\_$\rm K_S$& IRSF $\rm K_S$-band variability statistics (mag) \\
167-170 & Median\_Gaia\_G, ..., Amp\_Gaia\_G        & \textit{Gaia} G-band variability statistics (mag) \\
171-174 & Median\_Gaia\_BP, ..., Amp\_Gaia\_BP      & \textit{Gaia} BP-band variability statistics (mag) \\
175-178 & Median\_Gaia\_RP, ..., Amp\_Gaia\_RP      & \textit{Gaia} RP-band variability statistics (mag) \\
179-180 & Best\_Class\_Gaia, Best\_Class\_Score\_Gaia & \textit{Gaia} variable classifications \\
181-184 & Median\_NSC\_$u$, ..., Amp\_NSC\_$u$          & NSC $u$-band variability statistics (mag) \\
185-188 & Median\_NSC\_$g$, ..., Amp\_NSC\_$g$          & NSC $g$-band variability statistics (mag) \\
189-192 & Median\_NSC\_$r$, ..., Amp\_NSC\_$r$          & NSC $r$-band variability statistics (mag) \\
193-196 & Median\_NSC\_$i$, ..., Amp\_NSC\_$i$          & NSC $i$-band variability statistics (mag) \\
197-200 & Median\_NSC\_$z$, ..., Amp\_NSC\_$z$          & NSC $z$-band variability statistics (mag) \\
201-204 & Median\_NSC\_Y, ..., Amp\_NSC\_Y          & NSC Y-band variability statistics (mag) \\
205-214 & OGLE\_LPV\_Type, ..., OGLE\_LPV\_Iamp3    & OGLE long period variables statistics (mag) \\
215-241 & OGLE\_CCep\_Mode, ..., OGLE\_CCep\_phi31S & OGLE classical Cepheids statistics (mag) \\
242-252 & OGLE\_T2Cep\_Type, ..., OGLE\_T2Cep\_phi31& OGLE Type \uppercase\expandafter{\romannumeral2} Cepheids statistics (mag) \\
253-259 & OGLE\_Ecl\_Type, ..., OGLE\_Ecl\_DS       & OGLE eclipsing binaries statistics (mag) \\
\midrule  
\end{tabular}
\tablefoot{
This table is available in its entirety in CDS. \\
}
\end{table*}


\begin{table*}
\caption{Evolved Massive Star Candidate Selection Criteria} 
\label{outlines}
\begin{tabular}{ccc}
\toprule\toprule
Group & Color Criteria & Magnitude Criteria \\
\midrule
BSG$_{Gaia}$ & $(\rm BP-RP)<0.4$          & $\rm G\leq2.333\times(BP-RP)+14.617$  \\
             & $0.4\leq(\rm BP-RP)<0.65$  & $\rm G\leq-3.200\times(BP-RP)+16.830$ \\
YSG$_{Gaia}$ & $0.65\leq(\rm BP-RP)<1.1$  & $\rm G\leq14.750$                     \\
RSG$_{Gaia}$ & $1.1\leq(\rm BP-RP)<1.7$   & $\rm G\leq0.833\times(BP-RP)+13.833$  \\
             & $1.7\leq(\rm BP-RP)<2.2$   & $\rm G\leq-2.600\times(BP-RP)+19.670$ \\
             & $2.2\leq(\rm BP-RP)$       & $\rm G\leq13.950$                     \\
\midrule
BSG$_{SkyMapper}$ & $(r-i)<-0.05$        & $i\leq2.933\times(r-i)+15.797$  \\
                  & $-0.05\leq(r-i)<0.1$ & $i\leq-6.667\times(r-i)+15.317$ \\
YSG$_{SkyMapper}$ & $0.1\leq(r-i)<0.3$   & $i\leq14.650$                   \\
RSG$_{SkyMapper}$ & $0.3\leq(r-i)<0.65$  & $i\leq14.650$                   \\
                  & $0.65\leq(r-i)<0.9$  & $i\leq-5.200\times(r-i)+18.030$ \\
                  & $0.9\leq(r-i)$       & $i\leq13.350$                   \\
\midrule
BSG$_{NSC}$ & $(g-r)<0.05$        & $r\leq2.182\times(g-r)+15.541$  \\
            & $0.05\leq(g-r)<0.2$ & $r\leq-6.000\times(g-r)+15.950$ \\
YSG$_{NSC}$ & $0.2\leq(g-r)<0.7$  & $r\leq14.750$                   \\
RSG$_{NSC}$ & $0.7\leq(g-r)<1.3$  & $r\leq0.750\times(g-r)+14.225$  \\
            & $1.3\leq(g-r)<1.8$  & $r\leq-2.500\times(g-r)+18.450$ \\
            & $1.8\leq(g-r)$      & $r\leq13.950$                   \\
\midrule
BSG$_{M2002}$ & $(\rm V-R)<0.15$         & $\rm R\leq6.154\times(V-R)+14.527$  \\
              & $0.15\leq(\rm V-R)<0.25$ & $\rm R\leq-9.000\times(V-R)+16.800$ \\
YSG$_{M2002}$ & $0.25\leq(\rm V-R)<0.5$  & $\rm R\leq14.550$                   \\
RSG$_{M2002}$ & $0.5\leq(\rm V-R)<0.8$   & $\rm R\leq1.333\times(V-R)+13.883$  \\
              & $0.8\leq(\rm V-R)<1.1$   & $\rm R\leq4.000\times(V-R)+18.150$  \\
              & $1.1\leq(\rm V-R)$       & $\rm R\leq13.750$                   \\
\midrule
BSG$_{2MASS}$ & $(\rm J-K_S)<0.15$         & $\rm K_S\leq1.200\times(J-K_S)+14.670$  \\
              & $0.15\leq(\rm J-K_S)<0.3$ & $\rm K_S\leq-9.333\times(J-K_S)+16.250$  \\
YSG$_{2MASS}$ & $0.3\leq(\rm J-K_S)<0.5$  & $\rm K_S\leq-3.000\times(J-K_S)+14.350$  \\
RSG$_{2MASS}$ & $0.5\leq(\rm J-K_S)<1.0$   & $\rm K_S\leq-1.800\times(J-K_S)+13.750$ \\
              & $1.0\leq(\rm J-K_S)<1.3$   & $\rm K_S\leq-6.667\times(J-K_S)+18.617$ \\
              & $1.3\leq(\rm J-K_S)$       & $\rm K_S\leq9.950$                      \\
\midrule
\end{tabular}
\end{table*}

\begin{table*}
\caption{Numbers of Identified Evolved Massive Star Candidates}
\label{model_ident}
\begin{tabular}{ccccc}
\toprule\toprule
Dataset & Models & RSGs & YSGs & BSGs \\
\midrule
Gaia & MIST & 1029 & 97 & 952 \\
SkyMapper & MIST & 987 & 74 & 1035 \\
NSC & MIST & 627 & 34 & 606 \\
M2002 & MIST & 836 & 133 & 757 \\
2MASS & MIST & 1068 & 93 & 879 \\
2MASS & CB & 1020 & & \\
\midrule
Total (cleaned) & & 1405 & 217 & 1369 \\ 
\midrule
\end{tabular}
\end{table*}

\begin{table*}
\caption{Numbers of Ranked Evolved Massive Star Candidates}
\label{rank_number}
\begin{tabular}{cccc}
\toprule\toprule
Ranks & RSGs & YSGs & BSGs \\
\midrule
Rank 0 & 352 & 6  & 239 \\
Rank 1 & 421 & 17 & 362 \\
Rank 2 & 229 & 44 & 315 \\
Rank 3 & 105 & 51 & 188 \\
Rank 4 & 206 & 99 & 265 \\
Rank 5 & 92  &    &     \\
\midrule
Total  & 1405 & 217 & 1369 \\ 
\midrule
\end{tabular}
\end{table*}

\clearpage

\end{CJK*}

\end{document}